\documentclass [12pt, letterpaper]{article}
\pdfoutput = 1

\usepackage{amssymb,amsmath}
\usepackage{graphicx}
\usepackage{epsfig,amsfonts,amssymb}
\usepackage{amsmath}
\usepackage{amssymb}
\usepackage{mathrsfs}
\usepackage{cite}
\usepackage{mathtext}
\usepackage{epstopdf}
\usepackage{epsfig}
\usepackage{amsfonts}
\usepackage{psfrag}
\usepackage{amscd}
\usepackage{mathtext}
\usepackage[english]{babel}
\usepackage{setspace}
\usepackage{subfigure}
\usepackage{float}
\usepackage{fullpage}

\newcommand{\reef}[1]{(\ref{#1})}
\DeclareSymbolFont{AMSb}{U}{msb}{m}{n}
\DeclareMathSymbol{\IN}{\mathbin}{AMSb}{"4E}
\DeclareMathSymbol{\IZ}{\mathbin}{AMSb}{"5A}
\DeclareMathSymbol{\IR}{\mathbin}{AMSb}{"52}
\DeclareMathSymbol{\Q}{\mathbin}{AMSb}{"51}
\DeclareMathSymbol{\II}{\mathbin}{AMSb}{"49}
\DeclareMathSymbol{\IC}{\mathbin}{AMSb}{"43}
\DeclareMathSymbol{\IP}{\mathbin}{AMSb}{"50}
\DeclareMathSymbol{\IH}{\mathbin}{AMSb}{"48}
\DeclareMathSymbol\IA{\mathalpha}{AMSb}{"41}
\DeclareMathSymbol\IS{\mathalpha}{AMSb}{"53}

\def\Q{{\cal Q}}

\begin{document}

\begin{flushright}
\phantom{{\tt arXiv:1006.????}}
\end{flushright}

\bigskip
\bigskip
\bigskip

\begin{center}

{\Large \bf Non-perturbative String Theory from Water Waves}

\end{center}

\bigskip \bigskip \bigskip \bigskip

\centerline{\bf Ramakrishnan Iyer$^\natural$, Clifford V. Johnson$^\flat$, Jeffrey S. Pennington$^\sharp$}

\bigskip
\bigskip
\bigskip

\centerline{\it $^{\natural,\flat}$Department of Physics and Astronomy }
\centerline{\it University of
Southern California}
\centerline{\it Los Angeles, CA 90089-0484, U.S.A.}

\bigskip

\centerline{\it $^\sharp$SLAC National Accelerator Laboratory}
\centerline{\it Stanford University}
\centerline{\it Stanford, CA 94309, U.S.A.}

\bigskip

\centerline{\small \tt $^\natural$ramaiyer, $^\flat$johnson1,  [at] usc.edu; $^\sharp$jpennin [at] stanford.edu}

\bigskip
\bigskip

\begin{abstract}
\noindent
We use a combination of a 't Hooft limit and numerical methods to find
non--perturbative solutions of exactly solvable string theories,
showing that perturbative solutions in different asymptotic
regimes are connected by smooth interpolating functions. Our
earlier perturbative work showed that a large class of minimal string
theories arise as special limits of a Painlev\'e~IV hierarchy of
string equations that can be derived by a similarity reduction of the
dispersive water wave hierarchy of differential equations.  The
hierarchy of string equations contains new perturbative solutions,
some of which were conjectured to be the type~IIA and IIB string
theories coupled to $(4,4k-2)$ superconformal minimal models of type
$(A,D)$.  Our present paper shows that these new theories have smooth
non-perturbative extensions. We also find evidence for putative new
string theories that were not apparent in the perturbative analysis.
\end{abstract}
\newpage \baselineskip=18pt \setcounter{footnote}{0}

\section{Introduction}
In the quest to better understand non--perturbative phenomena in
string theory, and to find clues as to the full nature of
M--theory\cite{Hull:1994ys,Townsend:1995kk,Witten:1995ex}, rich
solvable examples are of considerable value. The type~0 minimal string
theories (formulated in
refs.~\cite{Morris:1990bw,Dalley:1991qg,Dalley:1991vr,Johnson:1992pu,Dalley:1992br}
and refs.~\cite{Crnkovic:1990ms,Hollowood:1991xq}, and recognized as
type~0 strings in ref.~\cite{Klebanov:2003wg}) have highly tractable
non--perturbative physics, and despite being exactly solvable contain a
rich set of phenomena such as holography and open--closed
dualities. Like the original minimal
strings\cite{Gross:1989vs,Brezin:1990rb,Douglas:1989ve}, they have
bosonic spacetime physics, although they have worldsheet fermions, and
may be thought of as two--dimensional supergravity coupled to ${\hat
  c}<1$ superconformal minimal models, with a diagonal GSO projection.

The physics of the minimal strings can be succinctly formulated in
terms of associated non--linear ordinary differential equations, known
as the string equations.  They furnish asymptotic expansions for the
free energy once the boundary conditions are fixed. These expansions
take the form of a string world--sheet expansion. Typically (for the
 cases of the type~0A and~0B systems coupled to the $(2,4k)$
superconformal minimal models) there are two perturbative regimes in
which such an expansion can be developed. One is interpreted as purely
closed string backgrounds with integer, $N$, units of R--R flux, while
the other has both open and closed string sectors with $N$ D--branes
in the background\footnote{Interestingly, in the associated
  non--linear system --- the generalized KdV system for type~0A ---
  the $N$ D--branes or flux units correspond to special soliton
  solutions, as shown in refs.~\cite{Carlisle:2005mk,
    Carlisle:2005wa}.}.

Remarkably, these models have a non--perturbative completion (first
discovered in
refs.~\cite{Morris:1990bw,Dalley:1991qg,Dalley:1991vr,Johnson:1992pu,Dalley:1992br}
in the context of the type~0A systems) that connects these two
asymptotic regimes, furnishing an example of a so--called ``geometric
transition'' between open and closed string backgrounds. While in the
case of type~0A, complete (numerical) solutions of the exact string
equations for any $N$ can be found, the connectedness of the type~0B
case has been argued for on the basis of a large $N$, 't Hooft--like,
limit where the interpolating solution can be found using algebraic
techniques\cite{Klebanov:2003wg}. This has more than just shades of
the AdS/CFT
correspondence\cite{Maldacena:1997re,Gubser:1998bc,Witten:1998qj},
where in the prototype, the open string physics of $N$ D3--branes can
be rewritten in terms of that of closed strings in AdS$_5\times S^5$
with $N$ units of R--R flux. In the present context we have a precise
analogue of this important example. There is of course a 't Hooft
large $N$ limit there too, connecting supergravity to large $N$
Yang--Mills.  That we have here not just the analogue of the large $N$
limit but also knowledge of how to go beyond, may ultimately prove
instructive.

In ref.~\cite{DWW}, we investigated a new system called the Dispersive
Water Wave (DWW) hierarchy\cite{Kaup1,Kaup2,Broer,Kuper,Gordoa:2001},
which we argued should yield new string equations {\it via} a
similarity reduction. These string equations turn out to be a
Painlev\'{e}~IV hierarchy of equations.  We found that both the
type~0A and~0B string theories were found to be naturally embedded
within this framework. In addition to this new non--perturbative
connection, we found further that the two string theories are merely
special points in a much larger tapestry of possibilities that also
appear to be string theories. This is somewhat suggestive of an
M--theory, now for minimal strings, which is exactly solvable. It
clearly deserves further study.

Among the other special points which suggested themselves (using
perturbation theory) were some that we conjectured to be the type IIA
and IIB string theories coupled to the $(4,4k-2)$ superconformal
minimal models. Much of the analysis carried out to support this
identification was using perturbative techniques.

The subject of the present paper is to continue the study of this rich
system into the non--perturbative regime. We argue that the
conjectured type~II theories have non--perturbative completions by
showing that the corresponding perturbative expansions match onto each
other smoothly. This is accomplished using a combination of analytical
and numerical techniques similar to those used for the type~0 theories
coupled to the $(2,4k)$ superconformal minimal models.

The outline of the paper is as follows: In sections 2 and 3, we review
essential aspects of the well--known type~0 theories and provide a
summary of results about the DWW system that will be needed later.  We
reproduce the analytic ('t Hooft limit) technique used for the type 0B
string theory, as in ref.~\cite{Klebanov:2003wg}, in section 4 to keep the
presentation self--contained.  We then reformulate it in a manner that
allows us to extend the technique to our DWW system. We apply this
strategy to our system in Section 5 for the first two even DWW flows,
which contain the type~0 and conjectured type~II theories of interest
and find that the type~II theories possess smooth non--perturbative
solutions. In addition, we find evidence for the existence of new
non--perturbative solutions with novel asymptotics.  In section 6, we
consider the first (and simplest) flow of the DWW system. For this
case, we are able to strengthen evidence for the existence of these
new non--perturbative solutions with exact numerical results. We end
with a brief discussion in section 7. An appendix details the
expansions arising from the DWW system for the first few flows along
with some of their essential properties.

\section{Type~0 strings: A brief review}\label{sec:type0review}
We begin with a brief review of the type~0 string theories coupled to the $(2,4k)$
superconformal minimal models of type $(A,A)$. Those familiar with
these models can proceed directly to the next section.
\subsection{Type 0A}\label{sec:type0Areview}
Type 0A string theory coupled to the $(2,4k)$ superconformal minimal
models (first derived and studied in
refs.\cite{Dalley:1991qg,Dalley:1991vr,Johnson:1992pu,Dalley:1992br}
and identified as type 0A in ref.~\cite{Klebanov:2003wg}) is described
by the string equation
\begin{equation}\label{streqn0A}
w\mathcal{R}^2 - \frac{1}{2}\mathcal{R}\mathcal{R}'' + \frac{1}{4} \mathcal{R}'^{2} = \nu^2 \Gamma^2 \quad ,
\end{equation}
where, for a particular model, $w(z)$ is a real function of the real
variable $z$, a prime denotes $\nu {\partial}/{\partial z}$, and
$\Gamma$ and $\nu$ are real constants. The quantity $\mathcal{R}$ is a
function of $w(z)$ and its $z$--derivatives. In general $w(z)$
additionally depends on couplings $t_k$ associated with flowing
between the various models. Then we have
\begin{equation}\label{GDpoly}
\mathcal{R} = \sum_{k = 0}^{\infty} \Big ( k + \frac{1}{2}\Big ) t_k P_k \quad ,
\end{equation}
where the $P_k[w]$ are polynomials in $w(z)$ and its $z$--derivatives,
called the {Gel'fand -- Dikii} polynomials\cite{Gelfand:1975rn}. They  are
related by a recursion relation (defining a recursion operator $R_2$)
\begin{equation}\label{KdVrec}
P'_{k+1} = \frac{1}{4}P'''_{k} - wP'_{k} - \frac{1}{2}w'P_{k}\equiv R_2\circ P'_k\ ,
\end{equation}
and fixed by the value of the constant $P_{0}$ and the requirement
that the rest vanish for vanishing $w$. The first four are:
\begin{eqnarray}\label{0APolys}
&&P_{0} = \frac{1}{2}\ , \quad P_{1} = -\frac{1}{4} w\ , \quad P_{2} = \frac{1}{16}(3 w^2 - w'')\ , \nonumber\\
{\rm and}\quad &&P_{3} = -\frac{1}{64}(10w^3 - 10ww'' - 5(w')^2 + w''')\ .
\end{eqnarray}
The $k$th model is chosen by setting all
the $t_j$ to zero except $t_{0} \equiv z$ and
\begin{equation}\label{t0A}
t_{k}=\frac{(-4)^{k+1}(k!)^2}{(2k+1)!}\ .
\end{equation}
This number is chosen so that the coefficient of $w^k$ in
$\mathcal{R}$ is set to $-1$.\footnote{This gives $w=z^{1/k} +\ldots$
  as $z\rightarrow+\infty$. If we had instead chosen $t_0=-z$, we
  would have chosen the coefficient of $w^k$ to be unity.}  The flows
between various models are captured by the integrable KdV
hierarchy\cite{Douglas, Banks}.

The function $w(z)$ defines the partition function $Z = \exp (-F)$ of
the string theory $via$
\begin{equation}\label{0AFreeEnergy}
w(z) = 2 \nu^2 \frac{\partial^2 F}{\partial \mu^2}\Big{|}_{\mu = z} \quad ,
\end{equation}
where $\mu$ is the coefficient of the lowest dimension operator in the
world--sheet theory.

The asymptotic expansions of the string equations for the first two
cases are:

\medskip

\noindent
{$k = 1$}
\begin{eqnarray}\label{0Aexpnsk=1}
w(z) &=& z + \frac{\nu \Gamma}{z^{1/2}} - \frac{\nu^2 \Gamma^2}{2 z^2} + \frac{5}{32} \frac{\nu^3}{z^{7/2}}\Gamma\left(4\Gamma^2 + 1\right) +\cdots  \quad (z \rightarrow \infty) \\
w(z) &=& 0 + \frac{\nu^2 (4 \Gamma^2 - 1)}{4 z^2} + \frac{\nu^4}{8} \frac{(4\Gamma^2 - 1)(4 \Gamma^2 - 9)}{z^5}+ \cdots \quad (z \rightarrow -\infty) \nonumber
\end{eqnarray}

\noindent
{$k = 2$}
\begin{eqnarray}\label{0Aexpnsk=2}
w(z) &=& z^{1/2} + \frac{\nu \Gamma}{2 z^{3/4}} - \frac{1}{24}\frac{\nu^2}{z^2}\left(6 \Gamma^2 + 1\right) + \cdots \quad (z \rightarrow \infty) \\
w(z) &=& (4 \Gamma^2 - 1)\left(\frac{\nu^2}{4 z^2} + \frac{1}{32}\frac{\nu^6}{z^7}(4 \Gamma^2 - 9)(4 \Gamma^2 - 25) + \cdots\right) \quad (z \rightarrow -\infty) \nonumber
\end{eqnarray}

The solutions for various $k$
for $z > 0$ can be numerically
and analytically shown to match onto the solution for $z<0$, providing
a unique\cite{Dalley:1991vr,Johnson:1992pu,Dalley:1992br}
non--perturbative completion of the theory. (See figure~\reef{fig:plot}
for an example of a solution found using numerical methods.)

\begin{figure}[ht]
  \begin{center}
  \includegraphics[width=100mm]{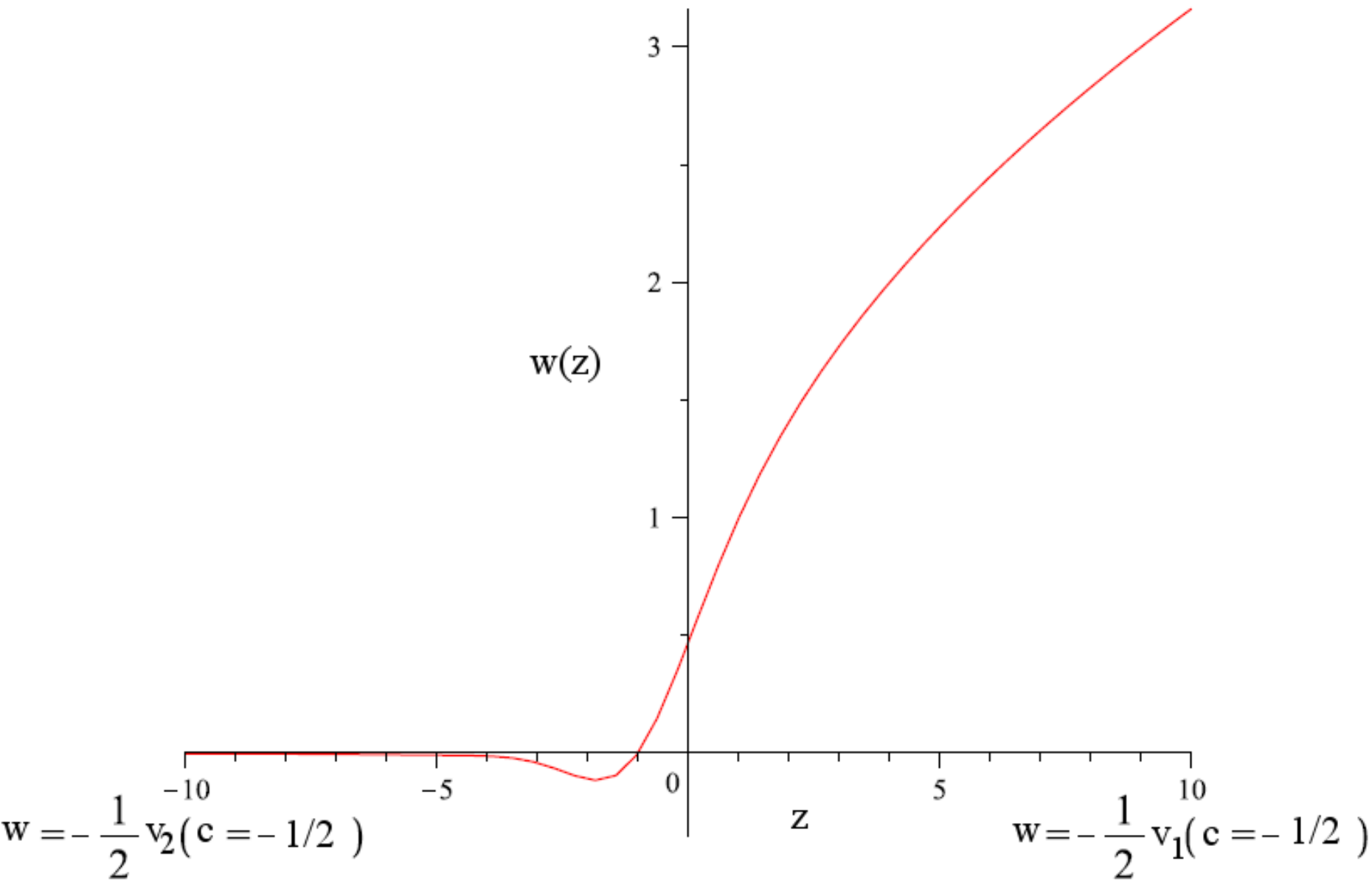}\\
  \caption{\footnotesize{A plot of the $k=2$ type~0A solution showing how the
    perturbative regimes at large $|z|$ are smoothly connected.
    Section~\ref{sec:DWWreview} discusses a function $v(x)$ ($x\propto z$),
    which has a number of different classes of behaviour distinguished
    by choice of boundary condition. The type~0A theory has class
    $v_1(z)$ in the $+z$ perturbative regime and class $v_2(z)$ in the
    $-z$ perturbative regime. Here we have set $\nu=1$ and
    $\Gamma=0$.}}\label{fig:plot}
  \end{center}
\end{figure}


In the $\mu \rightarrow +\infty$ regime, $\Gamma$
represents\cite{Dalley:1992br,Klebanov:2003wg} the number of
background ZZ D--branes~\cite{Zamolodchikov:2001ah} in the model, while in the
$\mu \rightarrow -\infty$ regime, $\Gamma$ represents the number of
units of R--R flux in the background\cite{Klebanov:2003wg}.

\subsection{Type 0B strings}\label{sec:type0Breview}
Type 0B string theory coupled to the $(2, 4k)$ superconformal minimal
models \cite{Klebanov:2003wg} is described by the following
{string} equations\cite{Crnkovic:1990ms,Hollowood:1991xq}:
\begin{eqnarray}\label{streqn0B}
\sum_{l = 0}^{\infty} t_{l}(l + 1)R_{l} = 0 \ ,\qquad
\sum_{l = 0}^{\infty}t_{l}(l + 1)H_{l} + \nu q = 0 \ ,
\end{eqnarray}
where the $R_l$ and $H_l$ are polynomials of functions $r(x)$ and
$\omega (x)$ (and their derivatives), and~$\nu$ and $q$ are real
constants.


The function ${\widetilde w}(x) = {r^2}/{4}$ defines the partition function of the
theory $via$
\begin{equation}\label{0BFreeEnergy}
 {\widetilde w}(x) = \frac{r^2}{4} =  \nu^2 \frac{d^2F}{dx^2}\Big{|}_{\mu = x}  \quad .
\end{equation}
where $\mu$ is the coefficient of the lowest dimension operator in the world--sheet theory.
The $n$th model is chosen by setting all $t_l$ to zero except $t_0
\sim x$ and $t_n$. These models have an interpretation as type 0B strings
coupled to the $(2,2n)$ superconformal minimal models only for even $n=2k$.
The flows between various models are captured by the integrable Zakharov--Shabat (ZS)
hierarchy\cite{Zakharov:1979zz}.


The asymptotic expansions of the string equations \reef{streqn0B} for
the first two even $n=2k$ are:

\noindent
$n = 2 \,\,\, (k=1)$
\begin{eqnarray}\label{0Bexpnsm=2}
{\widetilde w}(x) &=& \frac{x}{4} + \left(q^2 - \frac{1}{4}\right)\left[\frac{\nu^2}{2x^2} + \left(q^2 - \frac{9}{4}\right)\left(\frac{-2\nu^4}{x^5} + \cdots\right)\right]\ , \quad (x \rightarrow \infty) \nonumber\\
{\widetilde w}(x) &=&  \frac{\nu q\sqrt{2}}{4|x|^{1/2}} - \frac{\nu^2 q^2}{4 |x|^2} + \frac{\nu^3}{|x|^{7/2}}\frac{5\sqrt{2}}{64} q\left(1 + 4q^2\right)+\cdots  \quad (x \rightarrow -\infty)
\end{eqnarray}

\noindent
$n = 4 \,\,\, (k=2)$
\begin{eqnarray}\label{0Bexpnsm=4}
{\widetilde w}(x) &=& \frac{\sqrt{x}}{4} + \frac{\nu^2}{144 x^2} \left(64q^2 - 15\right) + \cdots ; \quad (x \rightarrow \infty) \\
{\widetilde w}(x) &=& \frac{\sqrt{|x|}}{2\sqrt{14}} + \frac{\nu}{2 |x|^{3/4}}\frac{q}{\sqrt{3}\cdot7^{1/4}} + \cdots  \quad (x \rightarrow -\infty) \nonumber
\end{eqnarray}


In the $\mu \rightarrow -\infty$ regime, $q$
represents the number of background ZZ D--branes in the model, while in
the $\mu \rightarrow \infty$ regime it counts the number of units of
R--R flux in the background\cite{Klebanov:2003wg}. The structure of
solutions with increasing $n$ is particularly rich \cite{Klebanov:2003wg}; the
$n=4$ expansion for $\mu \rightarrow -\infty$ shown above
breaks a $\IZ_2$ symmetry due to the presence of R--R fields
in the dual string theory.

Unlike the 0A case, the solutions for $x>0$ have not been shown
numerically to match onto those for $x < 0$ so far. The highly
non-linear nature of the string equations makes it difficult to
numerically obtain smooth solutions connecting the two regimes for the
0B case. Nevertheless, as argued in ref. \cite{Klebanov:2003wg} and
reviewed in section~\ref{sec:type0expnmtch}, these theories can be
argued to be non-perturbatively complete in a particular ('t Hooft)
limit.

For the $n=2$ ($k=1$) model\footnote{The central charge of the $\mathcal{N}=1$ $(p,q)$
super--conformal minimal models is given by $\hat{c} = 1 - \frac{2(p-q)^2}{pq}$.
For $n=2$, the central charge of the $(2,4)$ superconformal minimal model is thus zero
and we simply have the pure world--sheet supergravity
sector.}, the full non--perturbative solution is known since it can be mapped directly to
the solution known for the $k=1$ type~0A case $via$ the Morris map~\cite{Morris:1990bw,Morris:1992zr}.
The string equation for the 0A theory becomes the string equation for
the 0B theory once one identifies $\Gamma$ with $q$, but with the sign of
$x$ reversed. Analogues of the Morris map for higher $n$ are not known.

\section{The DWW system: A brief review}\label{sec:DWWreview}
The DWW system introduced in ref.~\cite{DWW} leads to a Painlev\'e~IV
hierarchy of string equations:
\begin{eqnarray}\label{DWWstring1}
-\frac{1}{2}\mathcal{L}_x+\frac{1}{2}u\mathcal{L}+\mathcal{K} &=& \nu c\quad \quad \quad \quad \quad  \\
\label{DWWstring2}
\left(-v+\frac{1}{4}u^2+\frac{1}{2}u_x\right)\mathcal{L}^2-\frac{1}{2}\mathcal{L}\mathcal{L}_{xx}+\frac{1}{4}\mathcal{L}_x^2&=&\nu^2 \Gamma^2\quad ,\quad \quad \quad \quad \quad
\end{eqnarray}
where $c$ and $\Gamma$ are real constants and $\nu$ plays the same role as
for the type~0 theories (note that here and in the rest of
the paper, for any function $G(x)$, $G_{x}$ will denote $\nu\,
\partial G/\partial x$).

The functions $u(x)$ and $v(x)$ are generalizations of the two
functions $r(x)$ and $\omega(x)$ used to describe the 0B theory. The polynomials
$\mathcal{L}[u,v]$ and $\mathcal{K}[u,v]$ are defined by
\begin{equation}\label{LK}
\left(\begin{array}{c}
                             \mathcal{L}\\
                             \mathcal{K}\end{array}\right) = \sum_{n=0}^{\infty}\frac{1}{2}(n+1)t_n \left(\begin{array}{c}
                             L_n[u,v]\\
                             K_n[u,v]\end{array}\right) \ \quad,
\end{equation}
where $L_n[u,v]$ and $K_n[u,v]$ are polynomials in $u(x)$, $v(x)$ and their
derivatives, similar to the polynomials $R_n$ and $H_n$ for the 0B theory.
They satisfy the recursion relation
\begin{eqnarray*}\label{DWWrec}
\left(\begin{array}{c}
             L_{n+1}[u,v]\\
             K_{n+1}[u,v]\end{array}\right) \ = R \left(\begin{array}{c}
                                                           L_n[u,v]\\
                                                           K_n[u,v]\end{array}\right) \ \quad .
\end{eqnarray*}
where $R$ is the recursion operator of the DWW hierarchy, given by
\begin{equation}\label{DWWRecOp}
R\equiv \frac{1}{2}\left(\begin{array}{cc}
                    \partial_{x}u\partial_{x}^{-1}-\partial_{x}&2\\
                    2v+v_{x}\partial_{x}^{-1}&u+\partial_{x}\end{array}\right) \quad .
\end{equation}

The first few $L_{n}$ and $K_{n}$ are as follows:
\begin{eqnarray}\label{DWWPolys}
L_0 &=& 2; \quad  K_0 = 0; \nonumber\\
L_1 &=& u; \quad  K_1 = v; \\
L_2 &=& \frac{1}{2}u^2 + v - \frac{1}{2}u_x; \quad   K_2 = uv + \frac{1}{2}v_x;\nonumber
\end{eqnarray}

The $n$th model is chosen by setting all $t_i$ equal to zero except for $t_0=x$ and $t_n$ which is chosen to be a numerical factor to fix the normalization. The parameter $t_n$ can be replaced by
\begin{equation}\label{tntogn}
 g_n \equiv \frac{1}{\frac{1}{2}(n+1)t_n} \quad ,
\end{equation}
in order to make direct contact with recent literature which discusses
this system in a much different (mathematical) context\cite{Gordoa:2001}.

Analytic expansion solutions to the string equations~\reef{DWWstring1} and~\reef{DWWstring2} for all $n$
can be organized into five main classes on the basis of
the $\nu^0$ (or $g_s^{-2}$) behavior of $u(x)$ and $v(x)$, as follows,
\begin{equation}\label{classes}
\begin{split}
\textrm{Class 1:}\quad&u_1 \sim 0\ ,\quad\quad\; v_1\sim x^{2/n}\ ;\\
\textrm{Class 2:}\quad&u_2 \sim 0\ ,\quad\quad\; v_2\sim 0\ ;\\
\textrm{Class 3:}\quad&u_3 \sim x^{1/n}\ ,\quad v_3\sim 0\ ;\\
\textrm{Class 4:}\quad&u_4 \sim x^{1/n}\ ,\quad v_4\sim x^{2/n},\quad u_4^2/v_4 \sim 1/4\ ;\\
\textrm{Class 5:}\quad&u_5 \sim x^{1/n}\ ,\quad v_5\sim x^{2/n},\quad u_5^2/v_5 \sim a\neq 1/4\ .\\
\end{split}
\end{equation}
They follow interesting patterns\footnote{For example, the expansions in Class 1 appear only for
even $n = 2k$, while the expansions in Class 5 are the analogues of the $\IZ_2$ symmetry-breaking
solutions of the 0B theories.} with increasing $n$ and have been explored in detail in
ref.~\cite{DWW}. Representative solutions for the first few $n$ have been provided in the Appendix
for ease of reference. In addition to the type~0 theories reviewed earlier, these solutions
encode new string theories, some of which were conjectured to be type~II string theories
coupled to super--conformal minimal models as reviewed below.

The type~0 theories reviewed earlier can be recovered completely from
this system of equations by applying appropriate
constraints~\cite{DWW}. Setting $u(x) = 0$ and $\mathcal{L} = 0$
in~\reef{DWWstring1} and~\reef{DWWstring2} results in the type~0A and
type~0B theories respectively\footnote{One also needs to make the
  identification $t_{2n}^{DWW}=
  \frac{1}{4}t_n^{\mathrm{KdV}}=\frac{(-1)^{n+1}4^n(n!)^2}{(2n+1)!}
  \; \Rightarrow \;g_{2n} =
  2\frac{(-1)^{n+1}(2n)!}{4^n(n!)^2}$.}. Consistency of the two
equations then requires that $c = -1/2$ for type~0A and $\Gamma = 0$
for type~0B, while the free parameter counts the branes and fluxes in
the respective theories. The function $v(x)$ (after appropriate
redefinition) encodes the free energies of the respective theories,
according to equations~\reef{0AFreeEnergy} and~\reef{0BFreeEnergy}.

One can also recover solutions corresponding to the type~0 theories by
setting $c=-1/2$ or $\Gamma = 0$ in appropriately chosen expansions
from the five classes listed above. Expansions in Classes 1 and 2
(labeled $v_1(x)$ and $v_2(x)$) reduce to those of the type~0A theory
(see~\reef{0Aexpnsk=1} and~\reef{0Aexpnsk=2}) once we set $c = -1/2$.
This helps fix the directions of the various expansions by requiring
that $v(x)$ be real, with the result that some expansions are real and
others complex for a given $n$. Further details can be found in
ref. ~\cite{DWW}.

A full solution for $v(x)$ is constructed by specifying its behavior
in the two asymptotic regimes, as seen for the $k=2$ 0A theory in
figure~\reef{fig:plot}. This 0A solution is obtained from the DWW
expansions, $v_1(x)$ and $v_2(x)$, by setting $c = -1/2$.  Using this
scheme, the above expansions can be organized in the form of a square
as shown in figure~\reef{square2}. The four corners of the square
correspond to four different string theories. The top corners
represent the known type 0 strings coupled to the $(2,4k)$ $(A,A)$
superminimal models.  The bottom corners represent two new theories
that were conjectured in ref.~\cite{DWW} to be type~IIA and IIB string
theories coupled to the $(4,4k-2)$ $(A,D)$ minimal models (for even
$k$ only). At each corner, one parameter from $(c, \Gamma)$ is frozen
to the value indicated on the figure, while the other parameter counts
the branes/fluxes in the appropriate regime.  The horizontal sides of
the square correspond to expansions in the $+x$ direction, while the
vertical sides represent expansions in the $-x$ direction. For all
$n$, there are two special points on the vertical sides corresponding
to $c =\pm\Gamma$, where the expansions $v_2$ and $v_3$ identically
vanish.

\begin{figure}[!ht]
  \begin{center}
 \includegraphics[width=80mm]{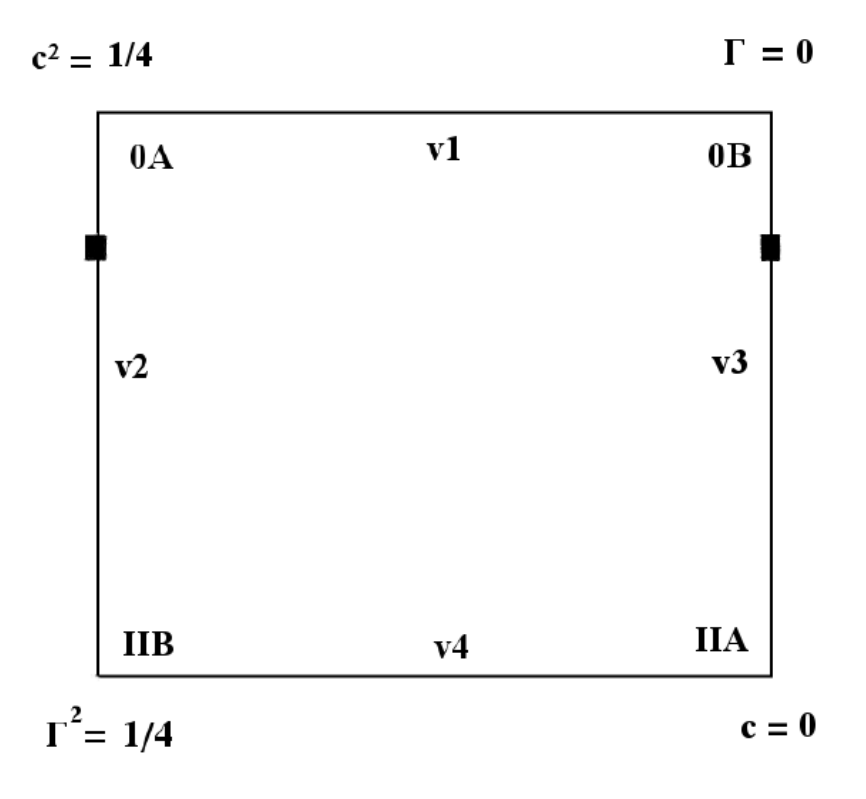}\\
 \caption{\footnotesize{A family of string theories, forming a square. See text for
   details.}}\label{square2}
 \end{center}
\end{figure}

 It must be noted that the square fully exists only
for even $n = 0\mod4$.  For $n=2\mod4$, the expansions labeled $v_4$
become complex as $+x$ expansions and the expansions labeled $v_1$ do
not appear for odd $n$.

The new theories were conjectured to be type~II strings coupled to super--minimal matter
by matching the genus zero contributions from the asymptotic expansions (the
terms appearing at order $g_s^0$ in the free energy) with the continuum calculation of
their one--loop partition functions. In the absence of non--perturbative constraints
(like $u = 0$ for the 0A theory) leading to the conjectured type~II theories,
the corresponding values of $c = 0$ and $\Gamma^2 = 1/4$ in figure~\ref{square2},
were obtained by systematic analysis of the properties of the genus zero
contributions~\cite{DWW}.

However, pairing perturbative expansions in this manner does not
guarantee the existence of a full non-perturbative solution with the
desired properties.  A type of 't Hooft limit will allow us to argue
for the existence of such non-perturbative solutions for the type~II
theories. We demonstrate how this limit works for the two parameter
DWW system in section~\reef{sec:DWWexpnmtch}.  Before that, we review
the 't Hooft limit for the known type~0 theories.

\section{Expansion Matching for type 0 theories : The known}
\label{sec:type0expnmtch}
The theories encoded by the string equations described so far can be
non--perturbatively completed by matching the perturbative solutions
for $x \rightarrow \infty$ on to those for $x \rightarrow
-\infty$. Numerical solutions have been found for the string equations
for the 0A models that smoothly interpolate between the two asymptotic
regimes as discussed earlier. See, e.g., figure~\reef{fig:plot}. The
string equations for the 0B models, however, are more difficult to
analyze numerically and attempts to find smooth numerical solutions
similar to the 0A ones have been unsuccessful thus far\footnote{The
  only exception is $n=2$ $(k=1)$ 0B theory, where the numerical $k=1$
  type 0A solution can be converted into it $via$ the Morris map, as
  mentioned in section~\ref{sec:type0Breview}.}.  This difficulty is
inherited by the DWW string equations presented in
section~\ref{sec:DWWreview}.

It was shown in ref.~\cite{Klebanov:2003wg} that the asymptotic expansions for the 0B theories
(for example, those in equations~\reef{0Bexpnsm=2} and~\reef{0Bexpnsm=4}) match on to each other
in a 't Hooft limit, with the brane/flux counting parameter (labeled $q$ for 0B and
$\Gamma$ for 0A) taken to be large. This was
then taken to suggest that the theories are non--perturbatively complete even when the
parameter is finite. We review this limit in some detail, before analyzing our system
in the same limit in section~\ref{sec:DWWexpnmtch}.

\subsection{The 0B theory in a 't Hooft limit}
\subsubsection{The $n=2$ theory}
The string equation for the simplest 0B theory with $n=2$ is
\begin{eqnarray}\label{streqn0Bm=2v2}
\nu^{2} \frac{\partial^2 r}{\partial x^2} -\frac{1}{2} r^3 + \frac{1}{2} x r + \nu^{2} \frac{q^2}{r^3} = 0 \quad ,
\end{eqnarray}
obtained by using the explicit forms of the polynomials $R_2$ and $H_2$
and eliminating $\omega(x)$ between the two equations~\reef{streqn0B}.

Consider the limit $q \rightarrow \infty$,  $x \rightarrow \pm \infty$ with
$t = (\nu q)^{-2/3}x$ fixed. This is our 't Hooft limit with $t^{-3/2}$ being the
't Hooft coupling\footnote{With
the identification $g_s = \frac{\nu}{x^{3/2}}$ and $q$ as counting the
number of units of R--R flux, this is recognized as the usual
't Hooft coupling $\lambda \sim g_s N$ familiar from higher-dimensional
string theories. Taking $x \rightarrow \pm \infty$ amounts to taking the limit
$g_s \rightarrow 0$.}. Defining $s = (\nu q)^{-1/3} r$, the string equation
\reef{streqn0Bm=2v2} becomes
\begin{equation}\label{0Bseqnm=2}
\frac{2}{q^2}s^3 \partial_{t}^2 s - s^6 + ts^4 + 2 = 0 \quad .
\end{equation}
In the large $q$ limit, the first term containing the derivative can be neglected
resulting in an algebraic equation for $f(t) = s^2$,
\begin{equation}\label{0Balgeqnm=2}
f^3 - t f^2 = 2 \quad .
\end{equation}
For generic $t$ this equation has 3 solutions, only \emph{one} of which is real
for all $t$ (an easy check is to consider $t\approx 0$). This solution reads
\begin{equation}\label{vsolnm=2}
f(t) = \frac{1}{3}\left[t + (t^3 + 27 - 3\sqrt{81 + 6 t^3})^{1/3} +
(t^3 + 27 + 3\sqrt{81 + 6 t^3})^{1/3}\right] \quad.
\end{equation}
For $t > -\frac{3}{2^{1/3}}$, the arguments of the square roots are positive and
$f(t)$ is real. Since there are opposite signs in front of the square
roots in the above expression, the half-integer powers of $t$ cancel when one
expands $f(t)$ for $t \rightarrow \infty$.

For $t < -\frac{3}{2^{1/3}}$ the arguments of the square roots are negative and the
second and third terms in the solution equation are complex, but $f(t)$ is real
since all the imaginary contributions cancel out.

The expansions for $f(t)$ for large negative and positive $t$ are
\begin{eqnarray}\label{0Bvexpnm=2}
f(t) &=& \frac{\sqrt{2}}{t^{1/2}} - \frac{1}{t^2} + \frac{5\sqrt{2}}{4t^{7/2}} -\frac{4}{t^5} + \cdots, \quad (t \rightarrow -\infty) \\
f(t) &=& t + \frac{2}{t^2} -\frac{8}{t^5} + \frac{56}{t^8} + \cdots \quad (t \rightarrow \infty) \nonumber
\end{eqnarray}
These expansions reproduce the coefficients of the highest powers of
$q$ in the asymptotic expansions to the string equation
\reef{0Bexpnsm=2} after remembering that $\widetilde{w} = f/4$. One
smooth function \reef{vsolnm=2} captures the limiting behavior of
large $q$ and connects\footnote{One can also formulate an equivalent
  argument as in ref.~\cite{Klebanov:2003wg} by starting with the
  action $S \sim \int dx \left[\frac{1}{2}r'^{2} + \frac{1}{8}(r^2 -
    x)^2 + \frac{1}{2}\frac{q^2}{r^2}\right] = \int dx
  \left[\frac{1}{2}r'^{2} + V(r^2)\right]$ whose equation of motion
  is~\reef{streqn0Bm=2v2}. Since this action is bounded below, it is
  clear that a solution to this equation will exist. The original
  matrix model integral dual to the theory is well defined and
  convergent, so one expects a finite and real answer for the free
  energy $F$. It is then natural to expect that
  equation~\reef{streqn0Bm=2v2} will have a unique real and smooth
  solution.} the two asymptotic regions smoothly.

This limit has been interpreted in~\cite{Klebanov:2003wg} as a 't
Hooft limit where only spherical topologies with boundaries
(D--branes) or fluxes survive. For $q = 0$, the 0B theory exhibits the
Gross--Witten phase transition at $x=0$ (namely $F'' = x/4 $ for $x >
0$ and $F'' = 0$ for $x < 0$). This transition can be smoothed out by
either the genus expansion with $q=0$ or by the expansion in the 't
Hooft parameter for large $q$.

\subsubsection{The $n=4$ theory}
A similar analysis shows that the asymptotic expansions for the $n=4$
theory \reef{0Bexpnsm=4} match smoothly onto each other. The analysis
is somewhat complicated because $\omega(x)$ cannot be eliminated in
favor of $r(x)$, unlike the $n=2$ case and one has to use the
asymptotic expansions for both $r(x)$ and $\omega(x)$. The expansions
for $\omega(x)$ are~\cite{Klebanov:2003wg}:
\begin{eqnarray}\label{0Bexpnsm=4omega}
\omega(x) &=& -\frac{2}{3}\frac{q}{x} + \frac{2}{3}\frac{q}{x^{7/2}}\left(\frac{80}{27}q^2 -\frac{5}{4}\right) + \cdots \quad (x \rightarrow \infty) \\
\omega(x) &=& -\frac{\sqrt{3}}{2}\left(\frac{2}{7}|x|\right)^{1/4} + \frac{2}{3}\frac{q}{|x|}
- \frac{5}{96\sqrt{3}}\left(\frac{7}{2}\right)^{1/4}\frac{3 + 32 q^2}{|x|^{9/4}} + \cdots \quad (x \rightarrow -\infty) \nonumber
\end{eqnarray}

In the $q \rightarrow \infty$ limit with $\omega = (\nu q)^{1/5} h$,
$x = (\nu q)^{4/5} t$ and $r^2 = (\nu q)^{2/5} f$, the derivative
terms in the string equation for this case are suppressed, resulting
in
\begin{eqnarray}\label{vheqnm=4}
\frac{3}{8}f^2 - 3 f h^2 + h^4 -\frac{3}{8}t &=& 0 \quad, \nonumber\\
f h \left(-\frac{3}{2}f + 2 h^2\right) &=& 1 \quad.
\end{eqnarray}
The second equation, on solving for $f$, gives
\begin{equation}\label{vpmeqnm=4}
f_{\pm} = \frac{2}{3}h^2 \pm \sqrt{\left(\frac{2}{3}h^2\right)^2 - \frac{2}{3h}} \quad .
\end{equation}
From equation~\reef{0Bexpnsm=4omega}, it can be seen that $h$ is negative, so that only the solution $f_{+}$
can be chosen\footnote{Note the presence of more than one solution for $f$. Such multiple solutions
will be prominent in our more general analysis in the next section.} for real $f$.

Substituting $f_+$ into the first equation gives, after some rearrangement:
\begin{equation}\label{heqnm=4}
-12 - 864 h^5 + 448 h^{10} - 36 ht - 96 h^{6} t - 27 h^{2} t^{2} = 0 \quad .
\end{equation}
Defining $y = ht$ reduces the above equation to a quadratic in $h^5$,
\begin{equation}\label{yeqnm=4}
- (864 + 96y)h^5 + 448h^{10} -(36y + 27y^2 + 12) = 0 \quad .
\end{equation}
Solving for $h^5$ as a function of $y$
\begin{equation}\label{hasfnofy}
h_{\pm}^5 = \frac{54 + 6y \pm 5\sqrt{3}\sqrt{28 + 3(y + 2)^2}}{56} \quad ,
\end{equation}
will allow a solution for $t = y/h$. The solution $h_{+}^5$ is always
positive and non--zero as a function of $y$. To get the asymptotics we
desire, we focus on $h_{-}^5$, which is negative and zero only for
$y_0 = -2/3$. Using $t = y/h_{-}$ it is easy to see that as $y
\rightarrow y_0$, $t \rightarrow \infty$ and and as $y \rightarrow
\infty$, $t \rightarrow -\infty$. As $t$ changes continuously, we will
always lie on the $h_{-}$ branch since the two branches never cross.
The expansion for $h_{-}$ as a function of $t$ can then be worked out
\cite{Klebanov:2003wg} to be
\begin{eqnarray}\label{0Bhexpnsm=4}
h_{-} &=& -\frac{\sqrt{3}}{2}(2 |t|/7)^{1/4} + \frac{2}{3|t|} - \frac{5(7/2)^{1/4}}{3\sqrt{3}|t|^{9/4}} + \cdots \quad (t \rightarrow -\infty)\nonumber\\
h_{-} &=& -\frac{2}{3t} + \frac{160}{81 t^{7/2}} + \cdots \quad (t \rightarrow \infty)
\end{eqnarray}
The coefficients in these expansions agree with the leading powers for
large $q$ in the expansions of $\omega(x)$ in
equation~\reef{0Bexpnsm=4omega} and can again be interpreted as a
limit in which only spherical topologies with boundaries or fluxes
survive.. This shows that an appropriately chosen solution of
equation~\reef{vheqnm=4} interpolates between the two asymptotic
regimes in the limit of large $q$.

This 't Hooft limit removes the derivative terms from the string
equations, resulting in algebraic equations that are simpler to
analyze. It can then be sensibly thought of as \emph{algebraic}
limit. Including the derivatives gives terms \emph{sub--leading} in
powers of $q$ and presumably does not introduce any singularities that
would destroy the smooth interpolation. The sub--leading powers of $q$
likely smooth out any sharp transitions (as we saw for the Gross--Witten phase
transition above) from large negative $x$ to large positive $x$.

\subsection{The 0A theory in a 't Hooft limit}
It is interesting to work out the 't Hooft limit of the type 0A
theories (with string equations~\reef{streqn0A}). Since, as already
discussed, solutions of the full equations have been obtained
numerically, it is instructive to compare the 't Hooft limit ({\it
  i.e.,} algebraic) results to the numerical results. (Note that while
the last section's type~0B results were a review, the type~0A analysis
is presented here for the first time.)

\subsubsection{The $k=1$ theory}
The string equation for this theory is,
\begin{eqnarray*}\label{streqn0Ak=1}
w\left(w-z \right)^2 - \frac{1}{2} \nu^2 \frac{\partial^2 w}{\partial z^2} \left(w-z\right) + \frac{1}{4} \nu^2 \left(\frac{\partial w}{\partial z} - 1\right)^2 = \nu^2 \Gamma^2 \quad .
\end{eqnarray*}
Consider the limit $\Gamma \rightarrow \infty$, $z \rightarrow \pm \infty$ with
$\rho = (\nu \Gamma)^{-2/3} z$ fixed. This is the 't Hooft limit with $\rho^{-3/2}$
being the 't Hooft coupling. Defining $w = (\nu \Gamma)^{-2/3} g$, the
above string equation becomes
\begin{equation}\label{0Aeqnforgk=1}
g (g - \rho)^2 - \frac{1}{2 \Gamma^2}(g - \rho)\partial_{\rho}^2 g
+ \frac{1}{4 \Gamma^2}(\partial_{\rho}g - 1)^2 = 1 \quad .
\end{equation}
In the large $\Gamma$ limit, the derivative terms can be neglected to give the simple
algebraic equation
\begin{equation}\label{0Aalgeqnforgk=1}
g (g - \rho)^2 = 1\  .
\end{equation}
As for the type~0B case of the last section, the solutions to this
cubic equation\footnote{One can also use the Morris map referred to at
  the end of section~\reef{sec:type0Breview} to obtain this equation
  directly from the corresponding 0B algebraic
  equation~\reef{0Balgeqnm=2}.} can be expanded as a Taylor series for
$\rho \rightarrow \pm \infty$ and seen to reproduce the coefficients
of the highest powers of $\Gamma$ in the asymptotic
expansions~\reef{0Aexpnsk=1}. One smooth function essentially connects
the two asymptotic regions, as in the 0B theory.

Figure~\reef{0Ak1g1} shows a comparison between the solution to
equation~\reef{0Aalgeqnforgk=1} and the  solution to the exact string
equations~\reef{streqn0Ak=1} obtained by numerical methods, both with
$\Gamma = 1$. The two solutions deviate slightly from each other in
the interior, and asymptotically, they match. This can be seen in
figure~\reef{0Ak1g1diff}, which plots the difference between the exact
and algebraic solutions, which goes to zero for large $|x|$. Physics
sub--leading in powers of $\Gamma$ contribute to the exact solution,
while only the highest powers of $\Gamma$ contribute to the algebraic
solution. The inclusion of the derivative terms does not introduce
any singularities.

\begin{figure}[!ht]
  \begin{center}
 \subfigure{
  \includegraphics[width=80mm, height = 60mm]{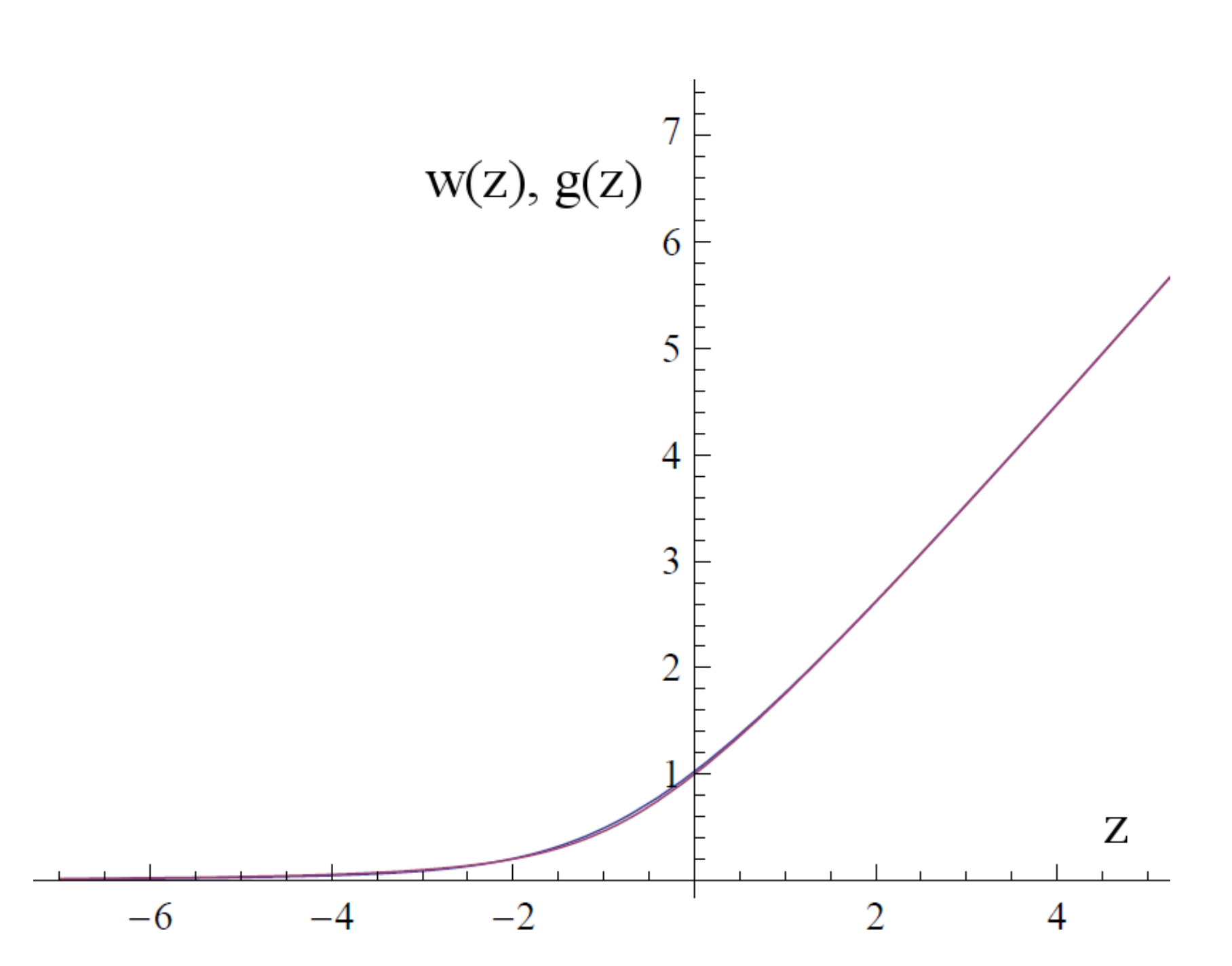}}
  \subfigure{\includegraphics[width=80mm, height = 60mm]{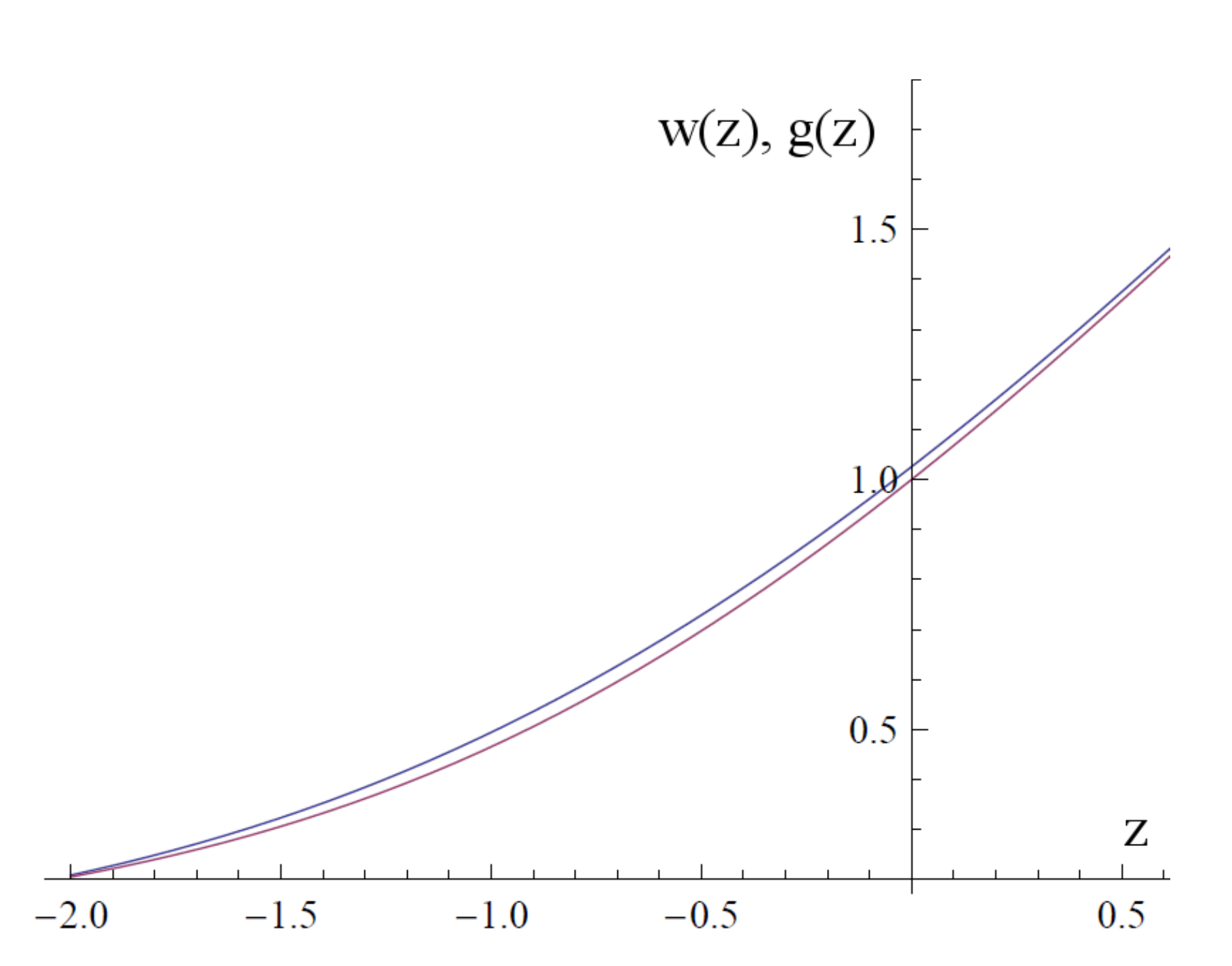}}
  \caption{\footnotesize{Comparison of the solution (in the 't Hooft
      limit) to equation~\reef{0Aalgeqnforgk=1} and the numerical
      solution to the full string equations~\reef{streqn0Ak=1}
      obtained by numerical methods, with $\Gamma = 1$. The curve
      which is uppermost at {\it e.g.,} $x=-1.0$ represents the
      solution of the exact equation. The plot on the right shows the
      same curves within a smaller domain for better
      resolution.}}\label{0Ak1g1}
 \end{center}
 \end{figure}

\begin{figure}[!ht]
  \begin{center}
 \includegraphics[width=80mm]{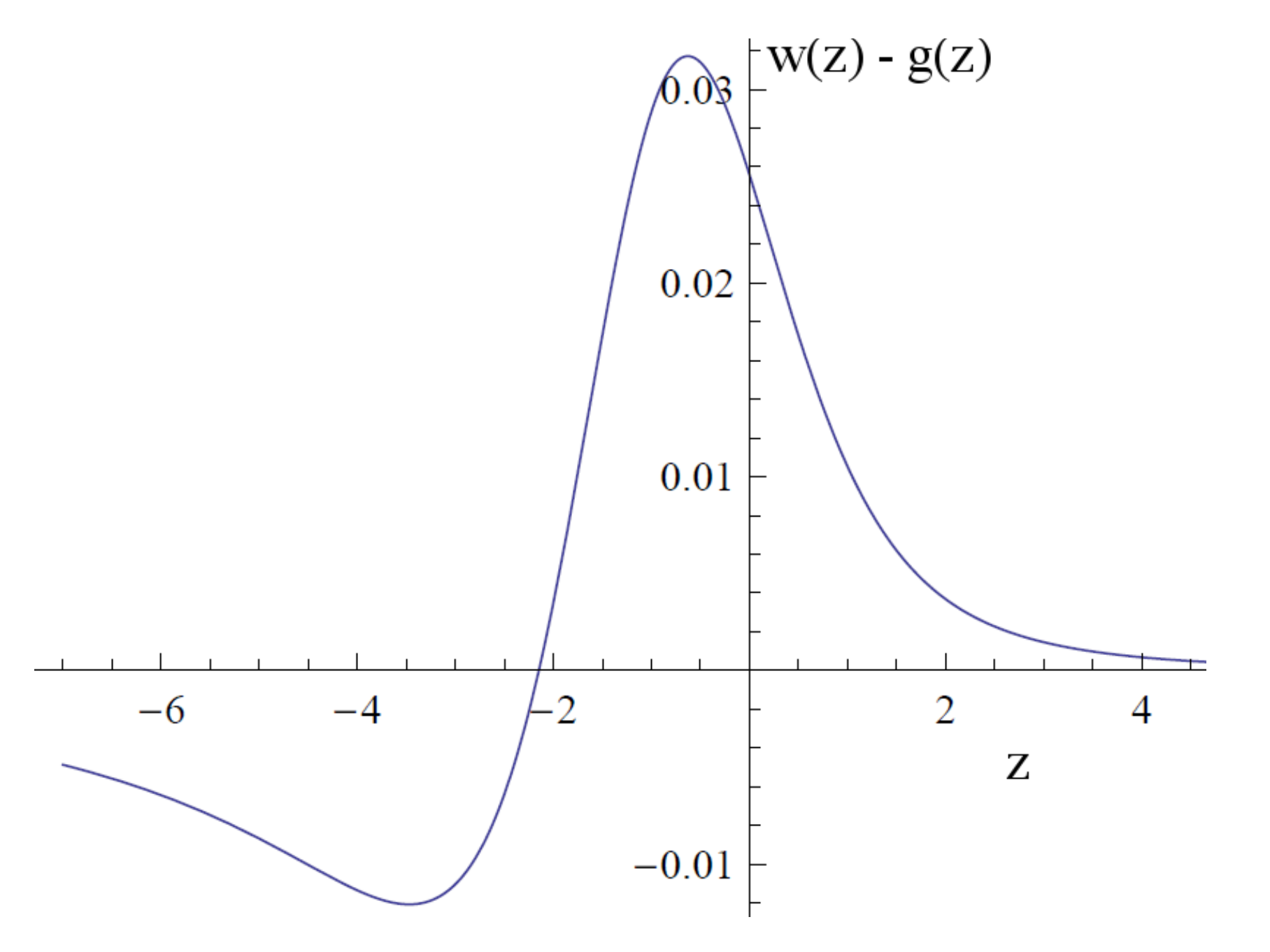}\\
 \caption{\footnotesize{The difference between the algebraic and full numerical solutions for the $k=1$
 0A theory. For large~$|x|$, the difference goes to zero.
   }}\label{0Ak1g1diff}
 \end{center}
\end{figure}

\subsubsection{The $k=2$ theory}
The string equation for the $k=2$ 0A theory has $\mathcal{R} = \frac{w^{''}}{3} - w^2 - z$. In
the large $\Gamma \rightarrow \infty$ and $z \rightarrow \pm \infty$ limit, with
$\rho = (\nu \Gamma)^{-5/4} z$ held fixed, the derivative terms drop out. Defining
$g = (\nu \Gamma)^{-5/4} w$ results in the algebraic equation
\begin{equation}\label{0Aalgeqnforgk=2}
g (g^2 - z)^2 = 1\ .
\end{equation}
The solutions to this equation reproduce the coefficients of the highest powers
of $\Gamma$ in the asymptotic expansions \reef{0Aexpnsk=2}.

\begin{figure}[!ht]
  \begin{center}
 \includegraphics[width=80mm]{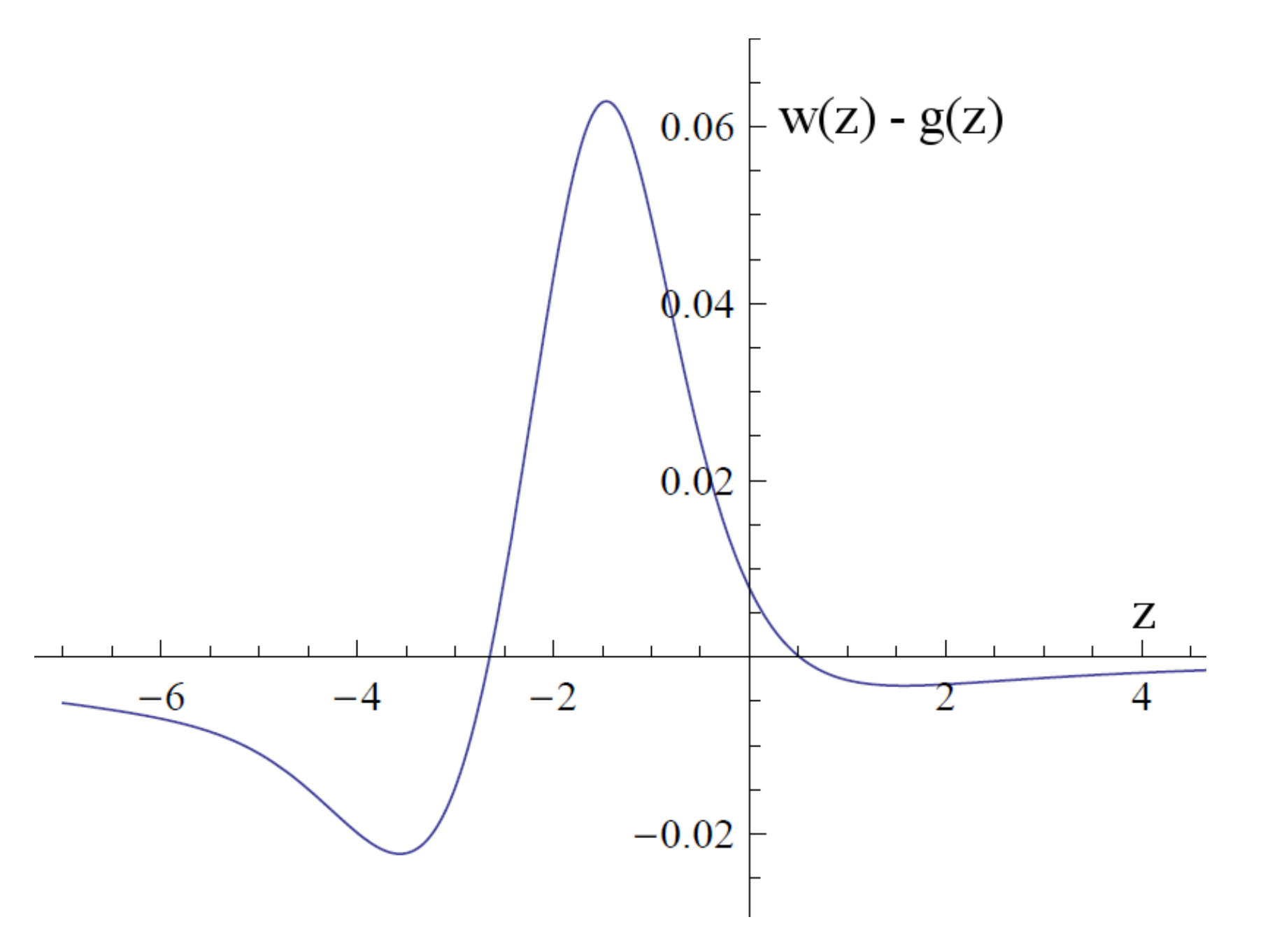}\\
 \caption{\footnotesize{The difference between the algebraic and full numerical solutions for the $k=2$
 0A theory. For large $|x|$, the difference goes to zero.}}\label{0Ak2g1diff}
 \end{center}
\end{figure}

Figure~\reef{0Ak2g1} shows a comparison between the solution to
equation~\reef{0Aalgeqnforgk=2} and the solution to the exact $k=2$ 0A
string equation obtained by numerical methods. The difference between
the two solutions is plotted in figure~\ref{0Ak2g1diff} and can be seen
to approach zero for large~$|x|$.  The deviation for finite $x$ and
asymptotic matching is evident, and can be explained by the additional
contributions coming from sub--leading powers of $\Gamma$ in the exact
solution.

\begin{figure}[!ht]
  \begin{center}
 \subfigure{
  \includegraphics[width=80mm, height = 60mm]{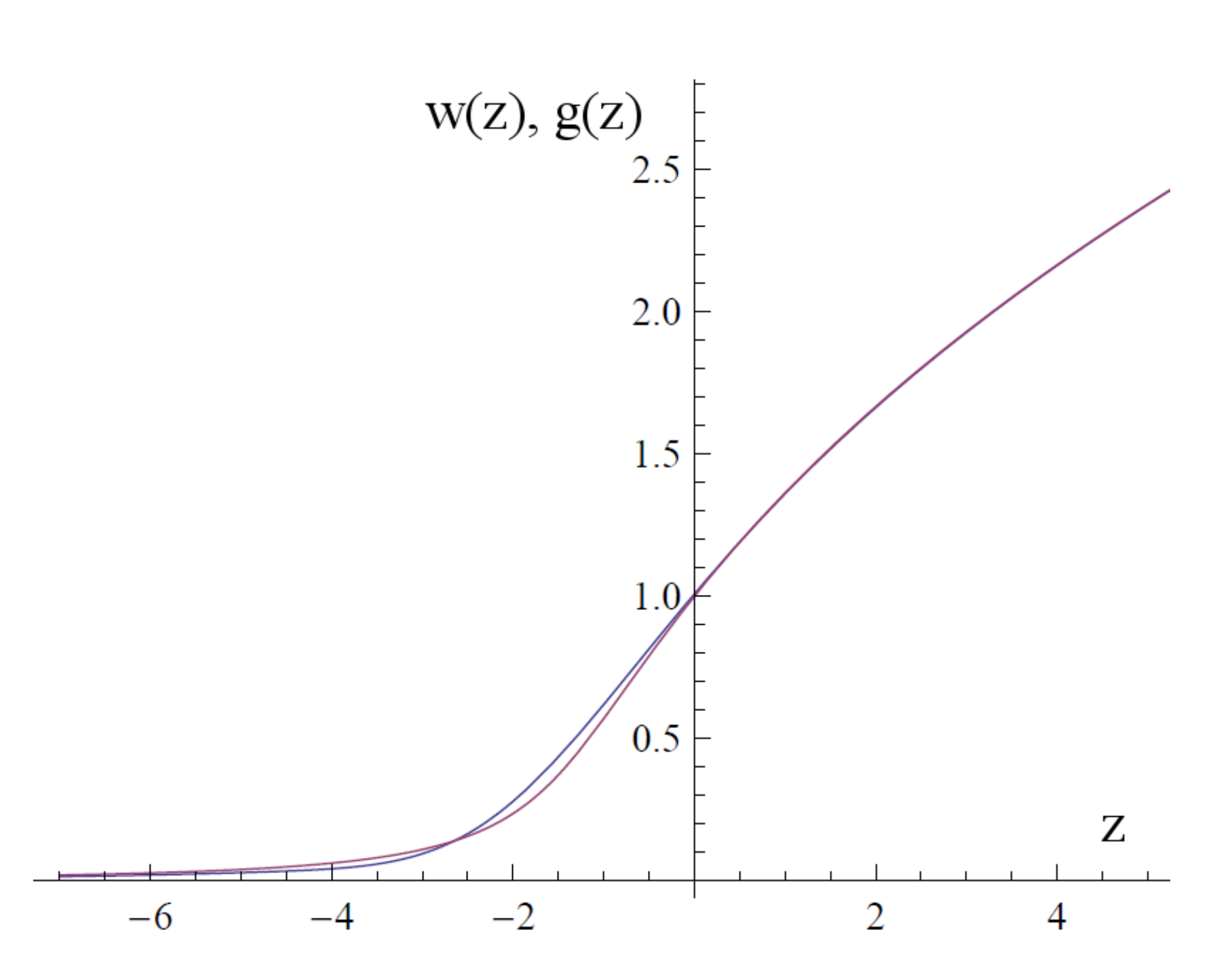}}
  \subfigure{\includegraphics[width=80mm, height = 60mm]{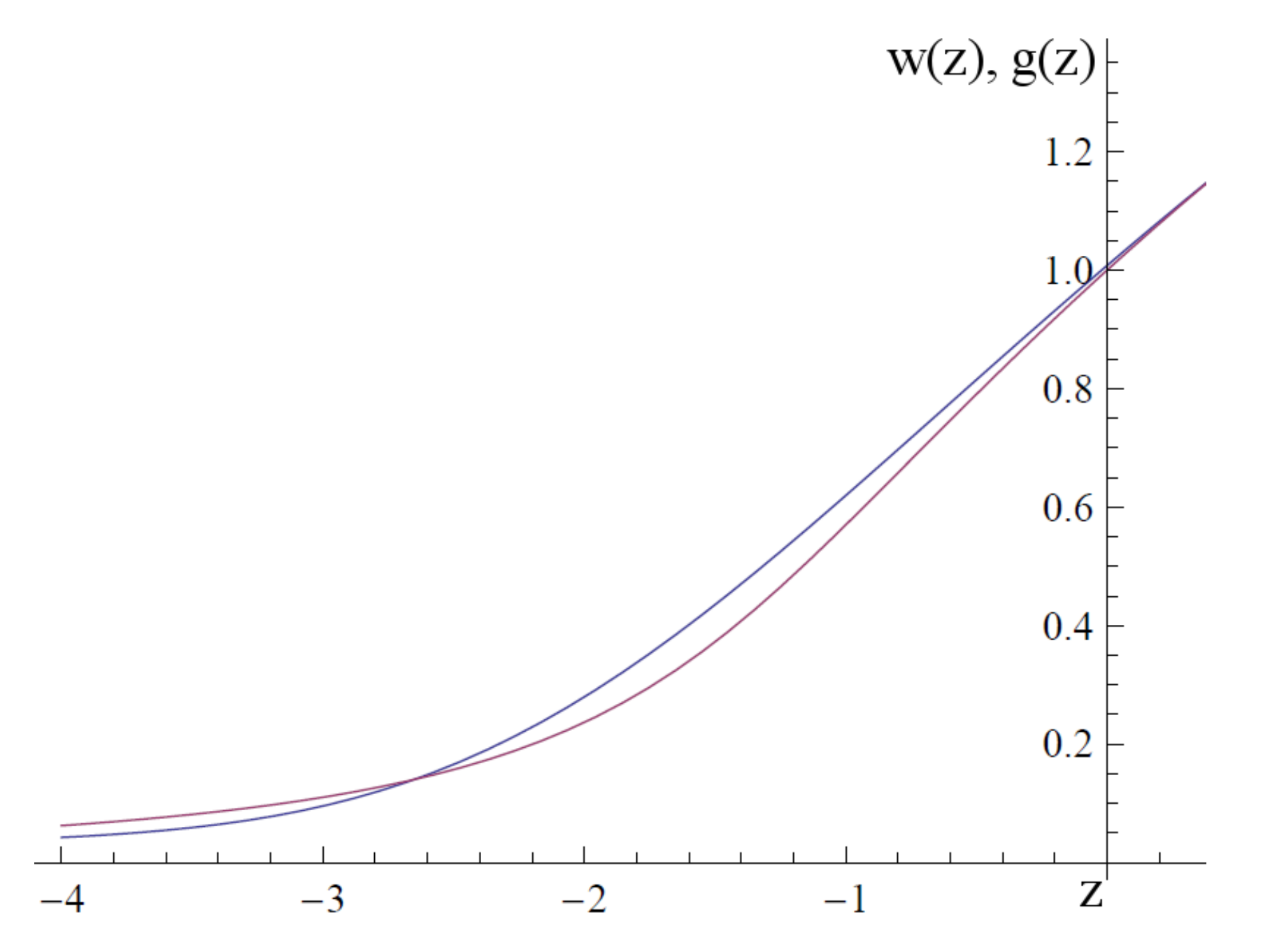}}
  \caption{\footnotesize{Comparison of the solution (in the 't Hooft
      limit) to equation~\reef{0Aalgeqnforgk=2} and the solution to
      the exact string equation 
      obtained by numerical methods, with $\Gamma = 1$. The curve
      which is uppermost at {\it e.g.,} $x=-2$ represents the
      solution of the exact equation. The plot on the right shows the same curves within a
      smaller domain for better resolution.}}\label{0Ak2g1}
 \end{center}
\end{figure}

\subsection{A modified 't Hooft limit}\label{sec:modlim}
A modification of the above limit (one that does not affect the
physics) will allow us to apply it to the Painlev\'e~IV hierarchy of
string equations. The need for such a modification will be explained
in the next section.

Instead of rescaling variables in the 0B (0A) theory to eliminate $q$ ($\Gamma$), one can
remove the derivatives by first rescaling
\begin{equation}\label{modlimit1}
q \rightarrow \frac{q}{\nu} \quad,
\end{equation}
and then taking the limit
\begin{equation}\label{modlimit2}
\nu \rightarrow 0 \quad,
\end{equation}
in that order, without affecting the physics. Note that the rescaled
$q$ is large, so this is still a large $q$ limit. The limit $\nu
\rightarrow 0$ amounts to taking $g_s \rightarrow 0$, as is clear from
the definition of~$g_s$ ($g_s = \frac{\nu}{\mu^{3/2}}$ for $n=2$ and 
$g_s = \frac{\nu}{\mu^{5/4}}$ for $n = 4$). This modified limit therefore extracts the
physics same as the 't Hooft limit of the previous subsections.

In this modified limit, the following equations are obtained
for the type 0B theories for $n=2$,
\begin{equation}\label{0Bralgeqnm=2}
r^6 - x r^4 = 2 q^2 \quad,
\end{equation}
and for $n=4$,
\begin{eqnarray}\label{0Balgeqnsm=4}
\frac{3}{8}r^5 - 3 r^3 \omega^2 + r\omega^4 - \frac{3}{8} x r &=& 0 , \nonumber\\
\frac{3}{2}r^4\omega - 2 r^2 \omega^3 &=& - q \quad .
\end{eqnarray}
These are the same as equations~\reef{0Balgeqnm=2} and~\reef{vheqnm=4}
with $r^2 \sim f$, $\omega \sim h$ and $x \sim t$, but with the rescaled parameter
$q$ explicitly present.

For the 0A theory, after rescaling $\Gamma \rightarrow \frac{\Gamma}{\nu}$, the
following equations are obtained for $k=1$ and $k=2$, respectively,
\begin{equation}\label{0Aalgeqnk=1}
w(w - z)^2 = \Gamma^2  ,\qquad
w(w^2 - z)^2 = \Gamma^2 \quad .
\end{equation}
In fact, for general $k$, the type~0A string equation \reef{streqn0A} effectively reduces to
\begin{equation}\label{0Aalgeqn}
w \widetilde {\cal{R}}^2 = \Gamma^2 \quad,
\end{equation}
where $\widetilde {\cal{R}}=w^k-z$ is obtained from $\cal{R}$ after
dropping all the derivatives of $w(z)$.

\section{Expansion Matching for DWW: The unknown}
\label{sec:DWWexpnmtch}
\subsection{\bf{'t Hooft limit for DWW : Two parameters}}
The string equations of the DWW hierarchy (reviewed in
section~\ref{sec:DWWreview}) are quite general, encoding the string
equations of type~0A and of type~0B as special cases. This generality,
however, comes at a price: all but the simplest cases are too
complicated to be solved using standard numerical methods. This
complicates our task of performing a non--perturbative analysis of the
string theories proposed in ref.~\cite{DWW}, which were based on
strictly perturbative considerations. To facilitate this analysis, we
examine the DWW string equations in a (modified) 't Hooft limit. The
result of this study is considerable evidence that the conjectured
type~II theories possess smooth solutions connecting the perturbative
expansions. This analysis also uncovers new non-perturbative
solutions, absent from the earlier perturbative analysis, which we
suspect may encode new unidentified string theories.

In section~\ref{sec:type0expnmtch} we described how to take the 't Hooft limit
of the type~0A and type~0B theories. Implicit in these methods was the fact
that the type~0 theories each depend on a single parameter, which counts branes or fluxes.
The full DWW equations, in contrast, have two free parameters, so it is not immediately
clear how to implement the 't Hooft limit in this more general case.

The perturbative analysis of ref.~\cite{DWW} suggests that, in
addition to the type~0 theories, the full DWW string equations also
describe type~II theories whose branes and fluxes are counted by a
single parameter. This motivates our assumption that \emph{new
  theories are described by a one--dimensional subspace of the
  parameters $c$ and $\Gamma$}. We further restrict our study to
subspaces defined by \emph{linear} combinations of $c$ and
$\Gamma$. This is the simplest possibility and although there could be
interesting one-dimensional subspaces corresponding to non-trivial
curves in $(c,\Gamma)$ parameter space, we will not consider them
here.

Given this assumption and the scaling prescription described in
section~\ref{sec:modlim}, we can utilize the 't Hooft limit to analyze
the string equations~\reef{DWWstring1} and~\reef{DWWstring2}.  For
given $\eta$ and $\xi$, let $c + \eta \Gamma = \xi$ be the constraint
defining the one-dimensional parameter subspace of a new theory under
consideration.  Since the constraint is (by assumption) satisfied by
the solutions of interest, we can impose the constraint on the DWW
equations themselves. We are left with a single free parameter which
we can recale as in section~\ref{sec:modlim} to take the 't Hooft
limit. In this way we obtain algebraic equations which are simple to
analyze.

Unfortunately, in the search for new theories, we do not \emph{a
  priori} know the values of~$\eta$ and $\xi$, so the above method is
not directly applicable. Nevertheless, we can proceed by adopting a
slightly different perspective. Note that the constraint $c + \eta
\Gamma = \xi$ implies that the free parameter can be taken to be $c$
or $\Gamma$, or any linear combination of them (as long as it is
independent of the constraint). Rescaling the free parameter by
$1/\nu$ is therefore equivalent to rescaling $c\rightarrow c/\nu$
and $\Gamma \rightarrow \Gamma/\nu$.  This procedure is independent of
the details of the original constraint, so it can be implemented
without actually knowing the values of $\eta$ and $\xi$. The result is
a set of algebraic equations which depends on $c$ and $\Gamma$.

Moreover, the solutions to these equations can be used to deduce
information about the original constraint. To see this, note that
since $\xi$ is a constant, after taking the limit the constraint takes
the form $c + \eta \Gamma = 0$. Therefore, the original constraint is
satisfied by any solution of the algebraic equations which satisfies
$c + \eta \Gamma = 0$. Turning this around, if a particular solution
to the algebraic equations only exists when $c + \eta \Gamma = 0$,
then this is a strong indication that it descends from the special
solution to the full equations subject to $c + \eta \Gamma =
\xi$. That is to say, it is an algebraic approximation to the full
solutions of some new string theory. Note that $\xi$ is not determined
from this approach, but $\eta$ is.

\subsection{'t Hooft limit for DWW : Strategy}
To obtain the algebraic equations, we begin with the DWW string
equations~\reef{DWWstring1} and~\reef{DWWstring2},
\begin{eqnarray*}
-\frac{1}{2}\mathcal{L}_x+\frac{1}{2}u\mathcal{L}+\mathcal{K} &=& \nu c\quad \quad \quad \quad \quad  \\
\left(-v+\frac{1}{4}u^2+\frac{1}{2}u_x\right)\mathcal{L}^2-\frac{1}{2}\mathcal{L}\mathcal{L}_{xx}+\frac{1}{4}\mathcal{L}_x^2&=&\nu^2 \Gamma^2\quad .\quad \quad \quad \quad \quad
\end{eqnarray*}
In the modified 't Hooft limit discussed above,
\begin{eqnarray}\label{modlimDWW}
\Gamma \rightarrow \frac{\Gamma}{\nu} \quad, \quad  c \rightarrow \frac{c}{\nu} \quad,\quad
\nu \rightarrow 0 \quad ,
\end{eqnarray}
the string equations simplify to
\begin{eqnarray}\label{DWWalgeqns}
\frac{1}{2}u \tilde{\mathcal{L}}+\tilde{\mathcal{K}} = c \ , \qquad {\rm and}\qquad
\left(-v+\frac{1}{4}u^2\right)\tilde{\mathcal{L}}^2 = \Gamma^2 \ ,
\end{eqnarray}
where $\tilde{\cal{L}}$ and $\tilde{\cal{K}}$ are the polynomials in $u$ and $v$
obtained from $\cal{L}$ and $\cal{K}$ after dropping all the derivative terms.
Before analyzing these equations in full generality, let us discuss a few special
cases to illustrate our methods.

\begin{enumerate}
\item $u = 0$ (type~0A)\\
The restriction to type~0A requires $n$ to be even, in which case
$\widetilde{\cal{K}} \propto u$. Thus, setting $u = 0$ in first equation of
\reef{DWWalgeqns} implies $c = 0$. This constraint can be written as $c +\eta \Gamma = 0 $
 with $\eta = 0$. As described above, this constraint lifts to $c+\eta \Gamma = c = \xi$
 at the level of the full equations. If we hadn't known the parameter constraint of
 the 0A theory ($c=-\frac{1}{2}$), the algebraic equations alone would indicate
 that $c=\xi$, a constant.

Alternatively one could start by looking for solutions to the equations~\reef{DWWalgeqns}
with $c = 0$. A subset of smooth solutions, identified by their
boundary behavior, will correspond to asymptotic expansions in Classes 1 and 2
with $c$ fixed (recall that these correspond to 0A). The boundary behavior of
the smooth solutions is matched with the behavior of expansions in Classes 1 and 2,
which are the asymptotic expansions of the type~0A theory.

\item $\widetilde{\cal{L}} = 0$ (type~0B)\\
Setting $\widetilde{\cal{L}} = 0$ forces $\Gamma = 0$ and the first equation
gives $\widetilde{\cal{K}} = c$. For even $n$, these
reduce to the 0B algebraic equations after redefining the variables
appropriately. The constraint $\Gamma = 0$ can be rewritten as
$c +\eta \Gamma = 0 $ with $\eta=\infty$ and lifts to a constraint
on the full equations which takes the form $\Gamma=\xi$, a constant.
The true parameter constraint for the full $0B$ theory is $\Gamma=0$;
the algebraic equations alone do not determine the precise value of $\xi$.

As in the 0A case above, one could look for smooth solutions to the algebraic
equations subject to $\Gamma = 0$. If they obey the correct boundary conditions (expansion
Classes 1 and 3 (or 5)) one could identify these as 0B solutions.

\end{enumerate}

Our strategy to demonstrate the existence of smooth non--perturbative solutions to
the type~0 and type~II theories and identify potentially new theories can be
summarized as follows. Take the limit~\reef{modlimDWW} of the full DWW equations
to obtain a set of algebraic equations. Search for solutions to these equations subject
to one of three constraints\footnote{It turns out that more general linear combinations
are relevant only for $n\geq 4$, and for simplicity we exclude such cases from the present
analysis.}:
\begin{equation}
c=0\ , \quad \Gamma = 0\ , \quad{\rm or}\quad c \pm \Gamma = 0\ .
\end{equation}
By matching the asymptotic behavior of these solutions onto the various
expansions, we identify the type~0 and type~II string theories.
We also find new solutions with asymptotics different from
those of the type~0 or type~II theories. We speculate that these
solutions correspond to some new unknown string theories.
As emphasized above, the precise parameter constraints of these new theories
cannot be fully determined from the algebraic equations.

\subsection{DWW $n = 2$ in a 't Hooft limit}
The string equations for $n=2$ reduce in the algebraic limit~\reef{modlimDWW} to
\begin{eqnarray}\label{n=2algeqn}
u\left( v + \frac{1}{2} u^2 + x\right) + 2 u v &=& 2 c \quad ,\nonumber\\
\left(u v - c\right)^2 - v \left(v + \frac{1}{2} u^2 + x\right)^2 &=& \Gamma^2 \quad .
\end{eqnarray}
They produce a total of nine asymptotic expansions which fall into 4
classes (see Appendix~\ref{Appexpns}).  As the expansions within each
class are related by various $\mathbb{Z}_2$ symmetries~\cite{DWW}, we
label each expansion only by a subscript specifying its class.  The
solutions corresponding to the three constraints in parameter space
are described below.
\begin{enumerate}
\item $c = 0$\\
  A plot of solutions to the algebraic equations~\reef{n=2algeqn} with
  $c = 0$ and $\Gamma=2$ is shown in figure~\reef{n=2c0}. The plot on
  the right shows two smooth solutions that connect
  regions\footnote{The algebraic solutions are represented by the
    solid black curves, while the asymptotic expansions are
    represented by the dashed curves. We adopt this convention in the
    rest of the paper.} of large $-x$ and $+x$. In our conventions,
  these are $v_2 | v_{1}$ and $v_{3} | v_{4}$ solutions\footnote{We
    label a solution with $-x$ asymptotics $v_L$ and $+x$ asymptotics
    $v_R$ by $v_L | v_R$.}. The plot on the left shows the solutions
  which do not join negative asymptotics to positive asymptotics, a
  feature we expect to find in any solutions describing new string
  theories. For this reason, these types of solutions will not be our
  primary interest.

\begin{figure}[ht]
  \begin{center}
  \subfigure{
  \includegraphics[width=78mm, height = 60mm]{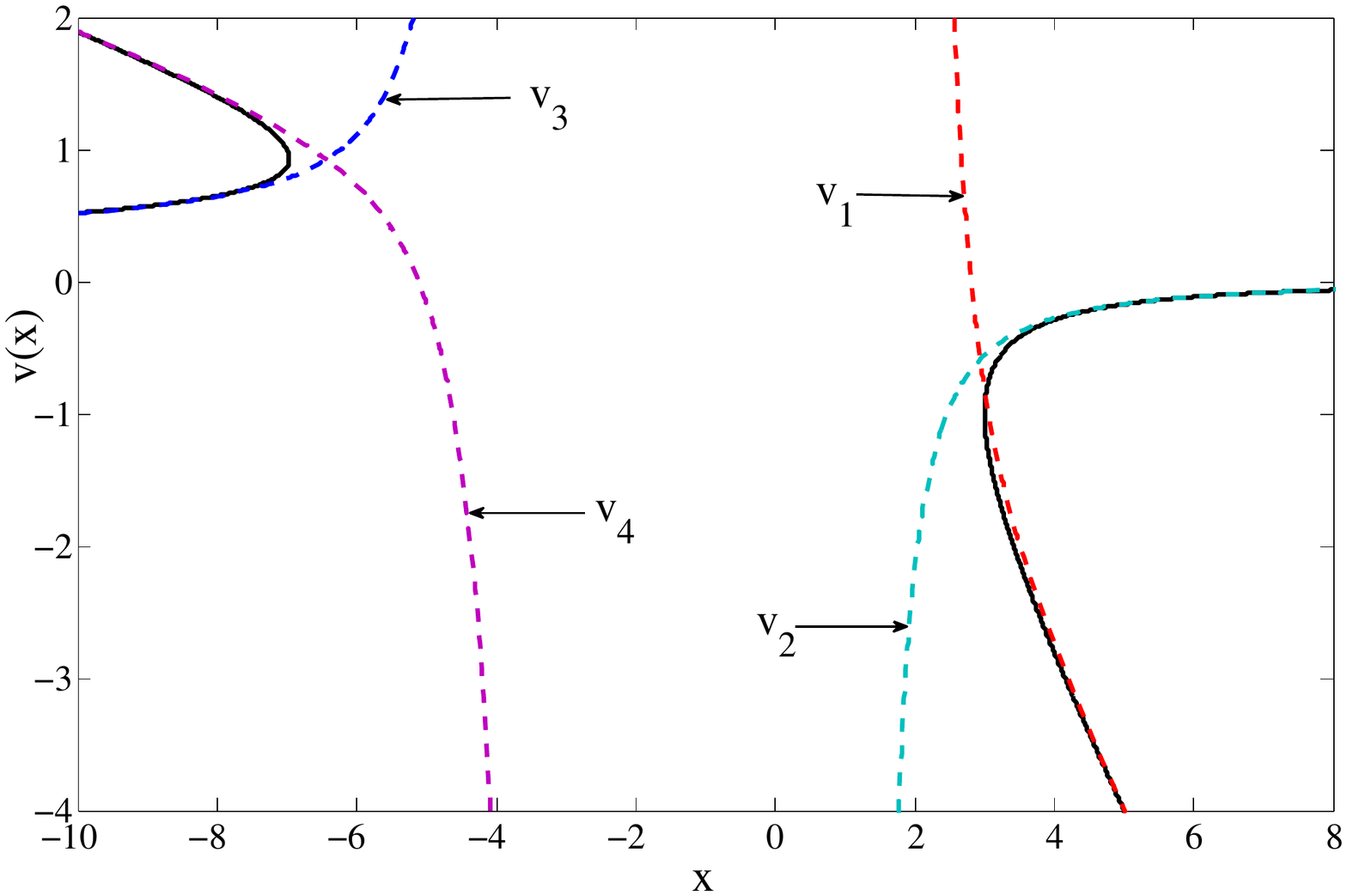}}
  \subfigure{\includegraphics[width=80mm, height = 60mm]{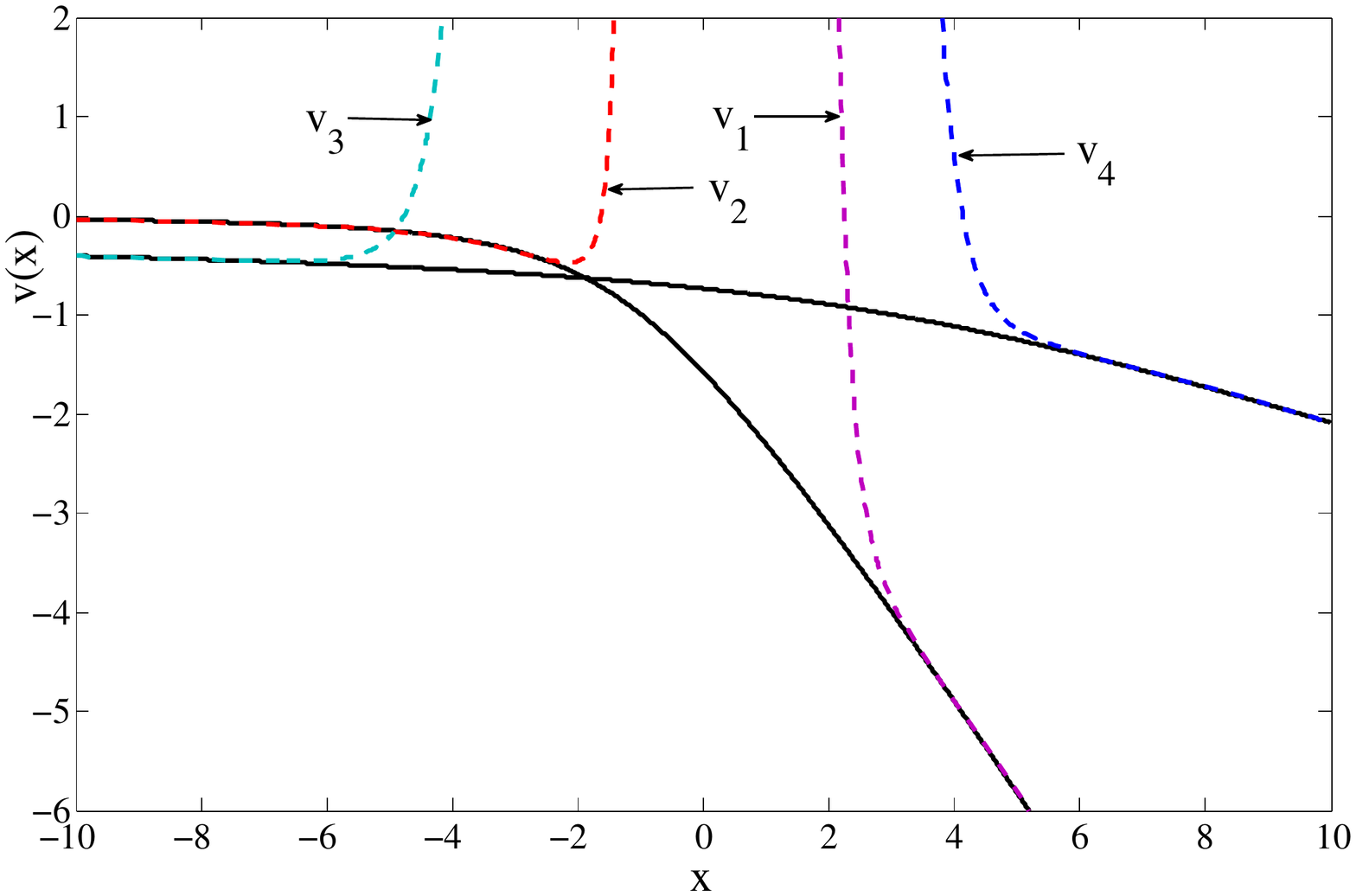}}
  \caption{\footnotesize{$n=2$ algebraic solutions with $c = 0$ and $\Gamma=2$.
  }}\label{n=2c0}
  \end{center}
\end{figure}

The $v_2|v_1$ solutions correspond to the 0A theory (with $k=1$)
coupled to pure supergravity.  It is unclear if the $v_3|v_4$
solutions correspond to a consistent new theory.  Based on the square
of theories in figure~\reef{square2}, it is tempting to conclude that
this theory might be the type~IIA string theory coupled to the $(4,2)$
$(A,D)$ superconformal minimal model. However, as mentioned in the
discussion below figure~\reef{square2}, the $v_4$ expansion is complex
as a $+x$ expansion for $n = 2\mod4$. For the particular case of $n=2$
and $c=0$, $v_4$ happens to be real, but this is an exception. We find
it unlikely that the corresponding solution to the full equations
exists and encodes a consistent theory, but we leave this question for
future investigation.

\item $\Gamma = 0$\\
Figure~\reef{n=2g0} shows the algebraic solutions for $\Gamma=0$ and $c=1.5$.
The right figure shows two smooth solutions, $v_{3}|v_{1}$ and $v_{4}|v_2$,
connecting the negative and positive regions.

The $v_{3}|v_{1}$ solutions are algebraic approximations to the full solutions of the
0B theory (with $n=2$) coupled to pure supergravity.
Analogous to the previous case with $c=0$, it is unclear if the $v_4|v_2$ solutions
descend from a consistent new theory. It is tempting to conclude that these are
type~IIB string theories coupled to the $(4,2)$ $(A,D)$ superminimal model, but this is likely
incorrect because $v_4$ is complex as a $+x$ expansion for $n=2\mod4$.

\begin{figure}[hh]
  \begin{center}
  \subfigure{
  \includegraphics[width=80mm, height = 59mm]{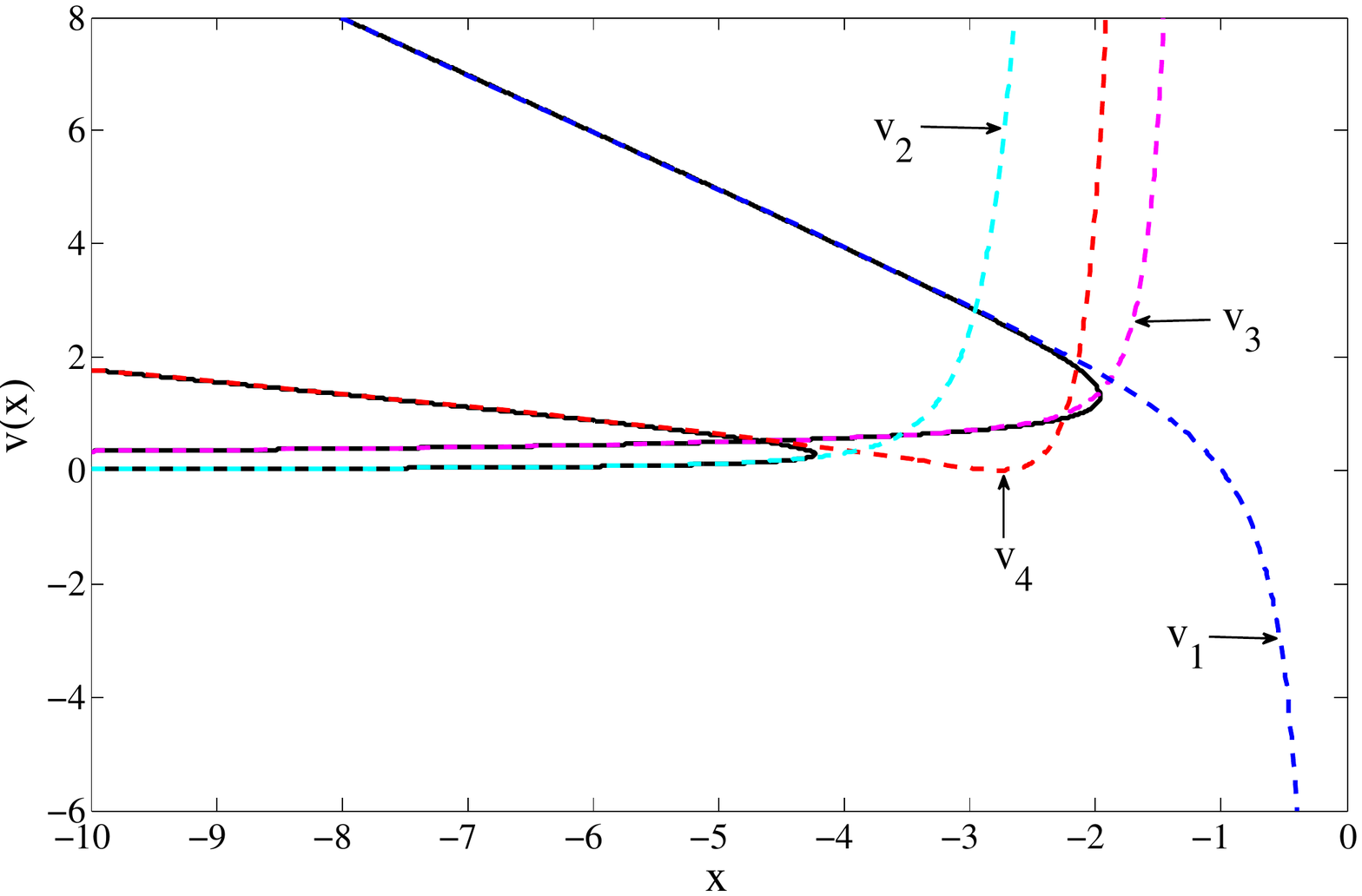}}
  \subfigure{\includegraphics[width=80mm, height = 60mm]{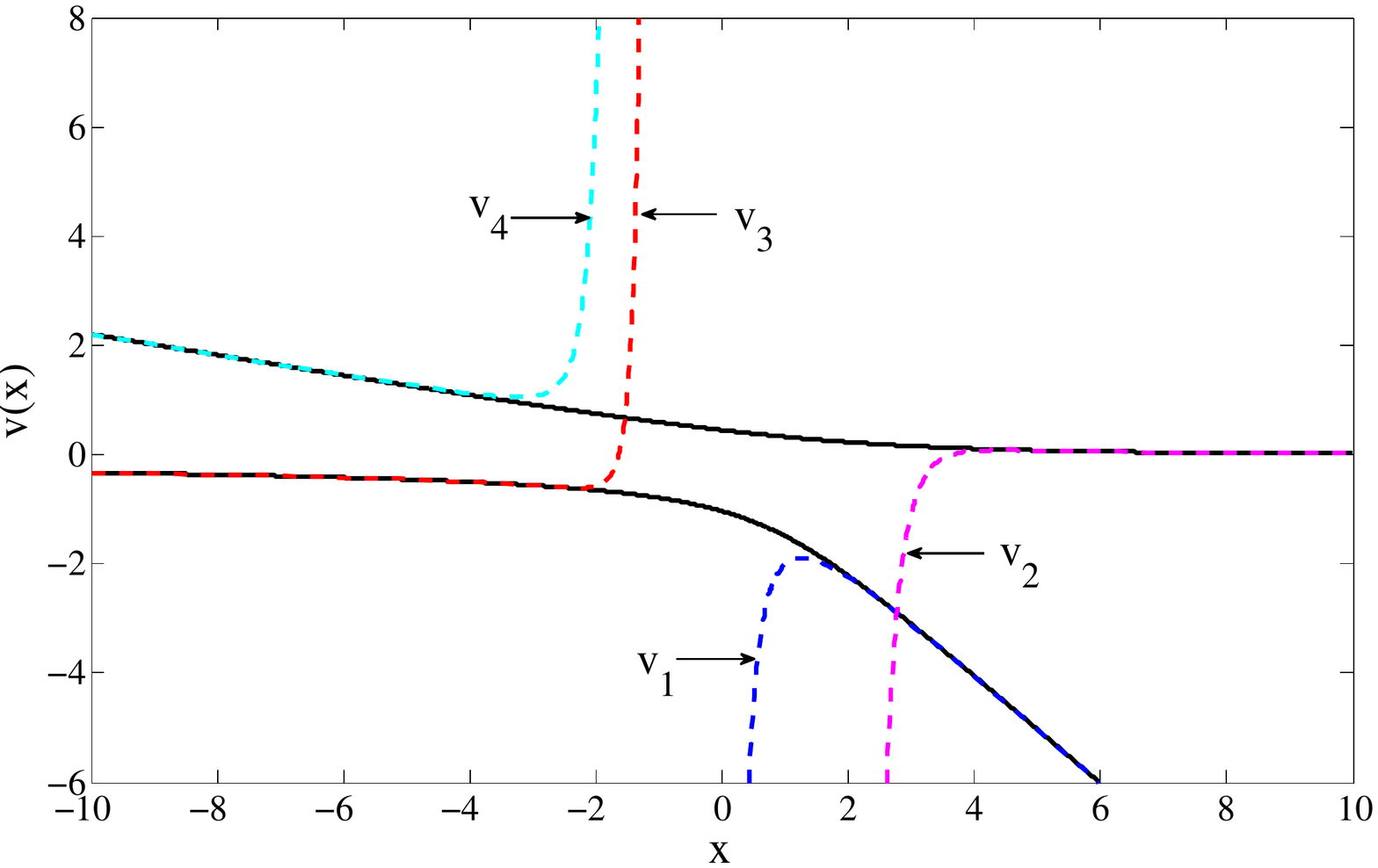}}
  \caption{\footnotesize{$n=2$ algebraic solutions with $\Gamma = 0$ and $c=1.5$.}
  }\label{n=2g0}
  \end{center}
\end{figure}

\item $c = \pm \Gamma$\\
We consider $c=\Gamma$. The case of $c=-\Gamma$ is completely analogous.
Figure~\reef{n=2ceqg} shows solutions for $c=2$ and $\Gamma=2$. The solutions
in the right figure are $v_4 | v_1$ and $v_3|v_1$ and connect the negative region
smoothly to the positive region.


\begin{figure}[ht]
  \begin{center}
  \subfigure{
  \includegraphics[width=80mm, height = 60mm]{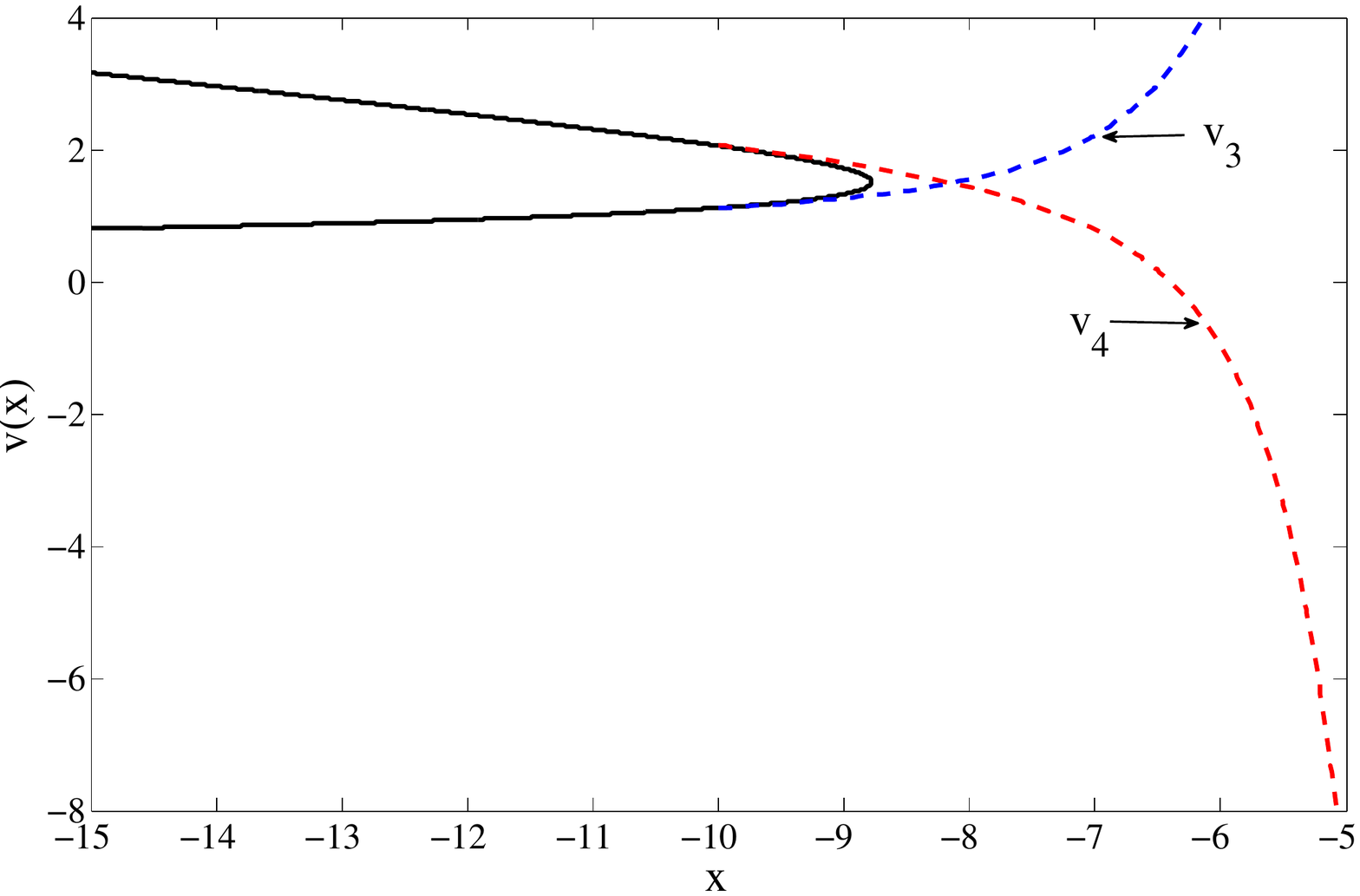}}
  \subfigure{\includegraphics[width=80mm, height = 60mm]{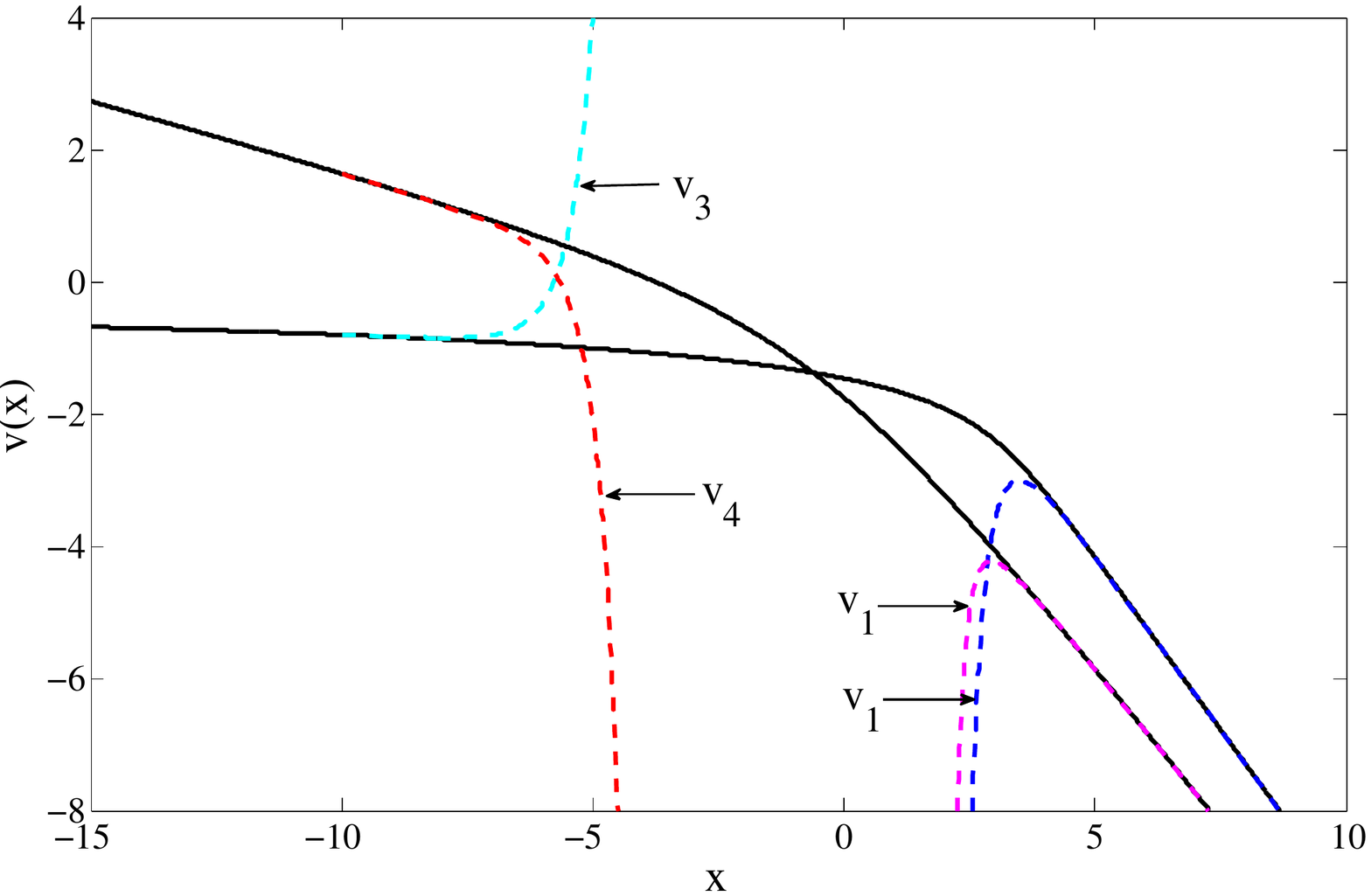}}
  \caption{\footnotesize{$n=2$ algebraic solutions with $c = 2$ and $\Gamma=2$.}
  }\label{n=2ceqg}
  \end{center}
\end{figure}

The fixed value of $c \mp \Gamma$ cannot be deduced from the algebraic
solutions.
\end{enumerate}

\subsection{DWW $n = 4$  in a 't Hooft limit}
The string equations in the 't Hooft limit, after setting $g_4 = -3/4$, are too long to
be reproduced here. The polynomials $\cal{L}$ and $\cal{K}$ reduce to,
\begin{eqnarray}\label{n=4algeqn}
\mathcal{\widetilde{L}} &=& \frac{3}{4}v^2 + \frac{3}{2} v u^2 + \frac{1}{8}u^4 - \frac{3}{4}x \quad, \nonumber\\
\mathcal{\widetilde{K}} &=& \frac{3}{2} u v^2 + \frac{1}{2} u^3 v \quad,
\end{eqnarray}
from which the corresponding algebraic string equations can be easily
obtained using~\reef{DWWalgeqns}.  There are a total of twenty--five
expansions~\cite{DWW}, falling into five classes. Solutions to the
algebraic string equations with the three constraints are shown below.

\begin{enumerate}
\item $c = 0$\\
Figure~\reef{n=4c0} shows solutions with $c=0$ and $\Gamma=2$. There are three solutions that
interpolate between the negative region and the positive region, $v_2|v_{1}$, $v_{5}|v_{3}$
and $v_{5}|v_{4}$.

\begin{figure}[ht]
  \begin{center}
  \subfigure{
  \includegraphics[width=80mm, height = 60mm]{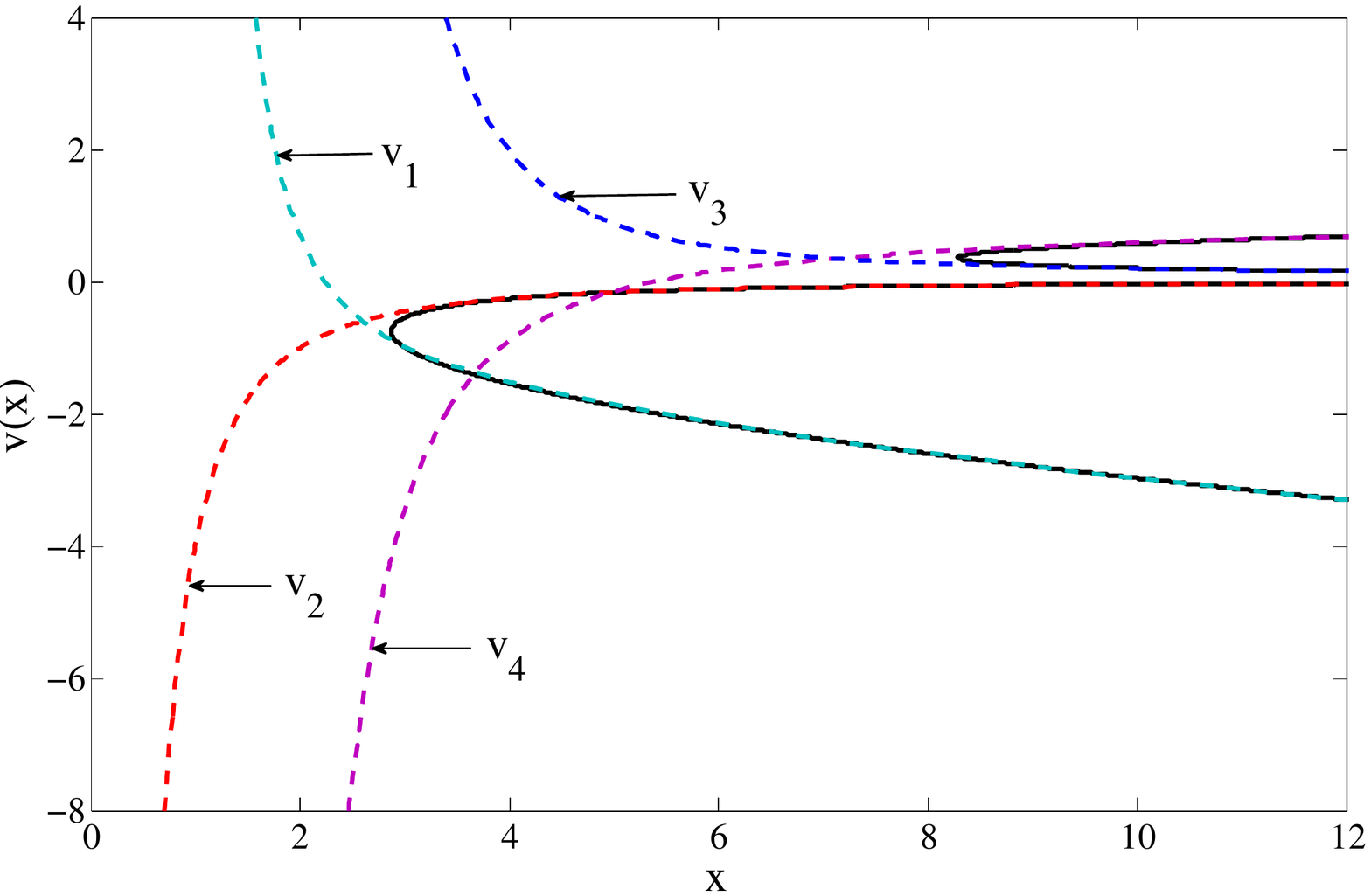}}
  \subfigure{\includegraphics[width=80mm, height = 60mm]{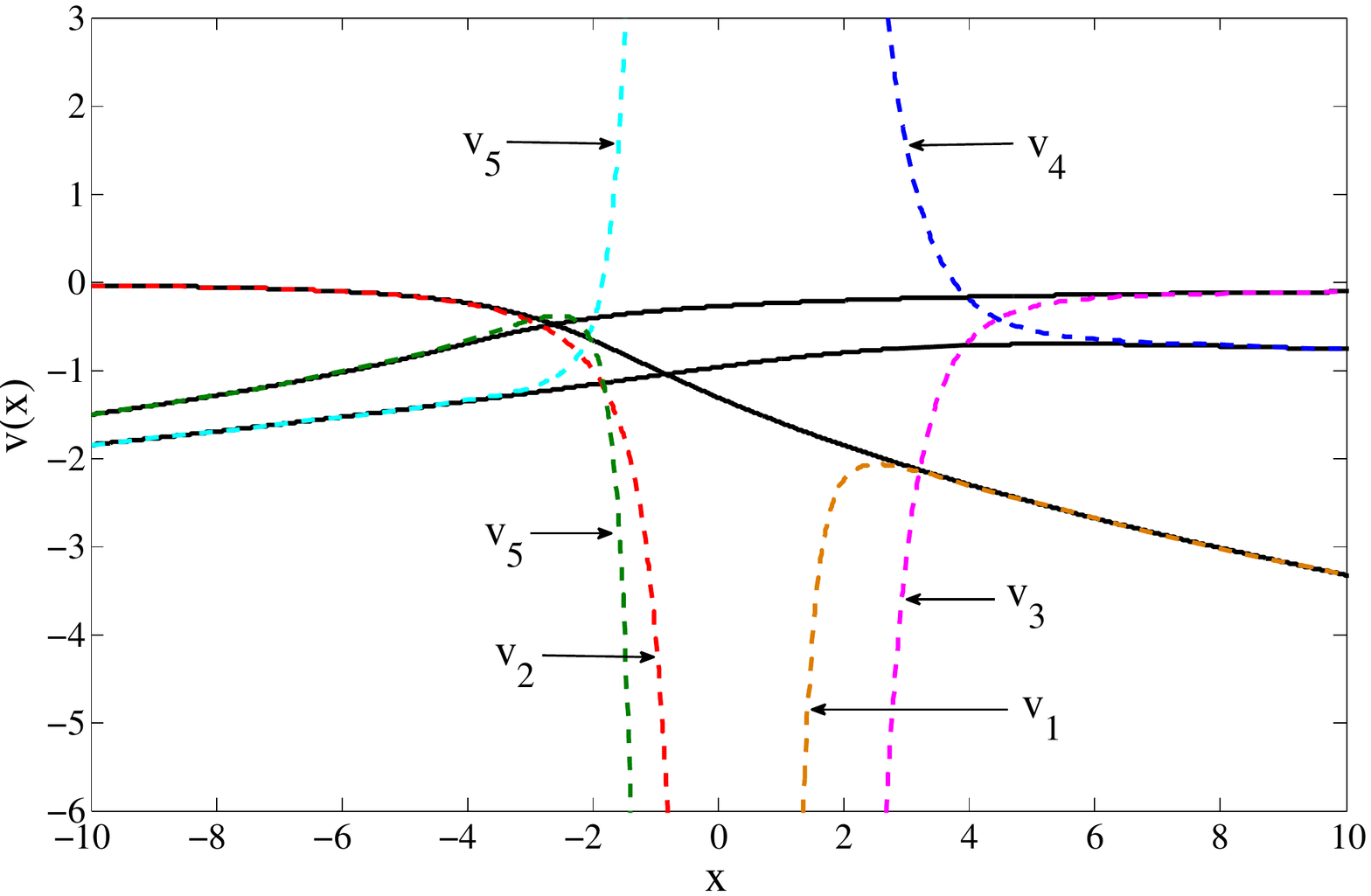}}
  \caption{\footnotesize{$n=4$ algebraic solutions with $c = 0$ and $\Gamma = 2$.}
  }\label{n=4c0}
  \end{center}
\end{figure}

The $v_2|v_1$ solutions represent type~0A coupled to the $(2,8)$
$(A,A)$ superminimal model. The $v_{5}|v_{4}$ solutions fit with our
conjectured type~IIA string theory coupled to the $(4,6)$ $(A,D)$
superminimal model and are analogues of the $\IZ_2$ symmetry--breaking
solutions of the 0B theory.

The $v_5 | v_3$ solutions are new.  On setting $c=0$ in the
\emph{full} expansions, the powers of $\Gamma$ in both correspond to a
parameter counting branes. An underlying string theory with these
solutions, if it exists, would have branes in both asymptotic
regimes\footnote{It is possible that such brane--brane solutions could
  be summed up to form rational solutions. This would be analogous to
  the rational solutions of the type~0A string equations that were
  considered in a string theory context in
  ref.\cite{Johnson:2006ux}. The rational solutions have $v_2$ type
  expansions (for $c=-1/2$) in both asymptotic directions for
  $x$.}. Further work is needed to conclude if such an underlying
theory exists.

\item $\Gamma = 0$\\
  Figure~\reef{n=4g0} shows solutions with $\Gamma=0$ and $c=1$. The
  solutions which interpolate between $-x$ and $+x$ are $v_{5}|v_{1}$,
  $v_{5}|v_{3}$ and $v_2 | v_{4}$. The $v_5|v_1$ solutions correspond
  to the $\IZ_2$ symmetry--breaking solutions of type~0B string theory
  coupled to the $(2,8)$ superminimal model, while the $v_2|v_4$
  solutions fit with our conjectured type~IIB string theory coupled to
  the $(4,6)$ $(A,D)$ superminimal models.

\begin{figure}[ht]
  \begin{center}
  \subfigure{
  \includegraphics[width=80mm, height = 60mm]{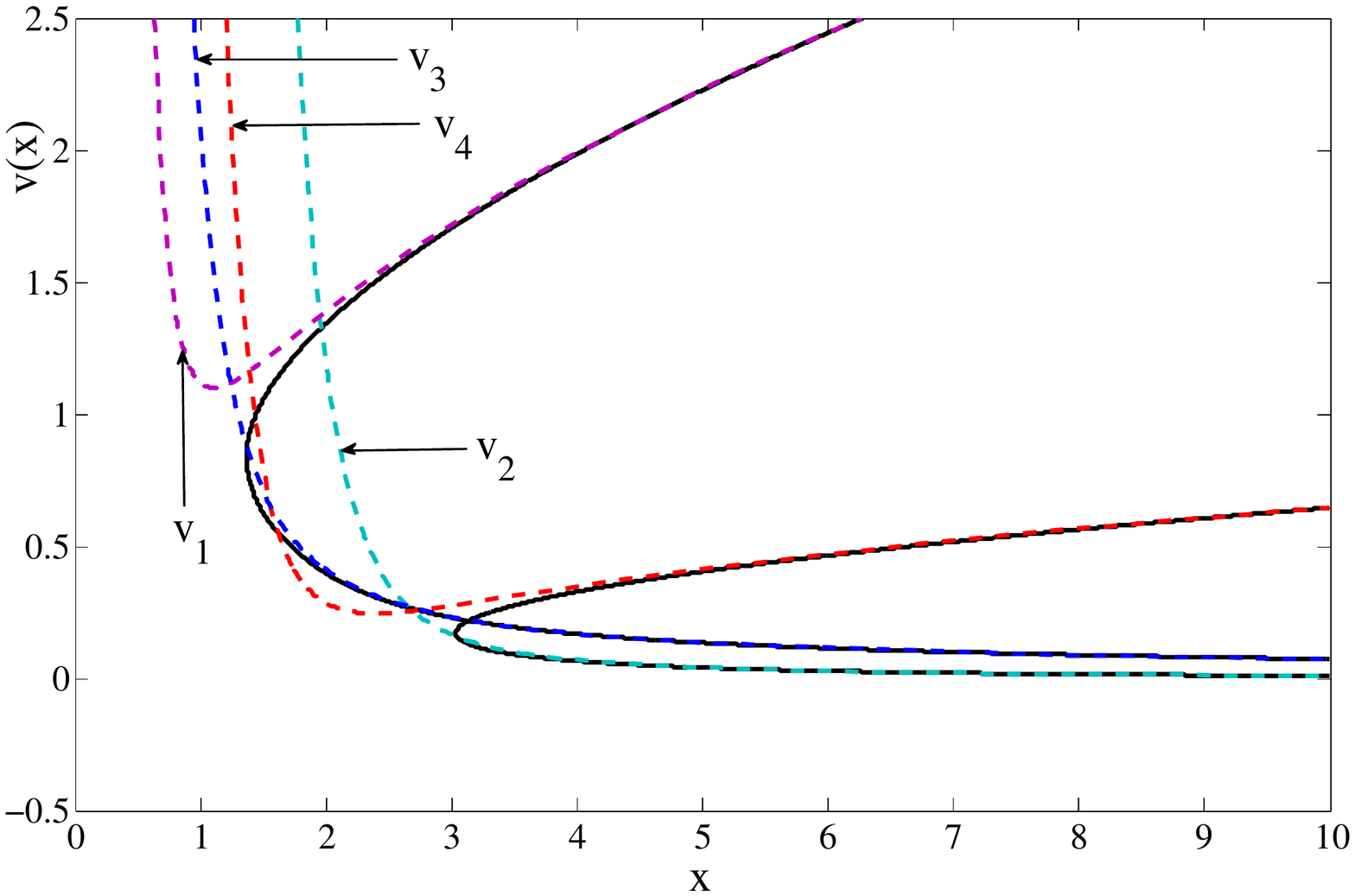}}
  \subfigure{\includegraphics[width=80mm, height = 60mm]{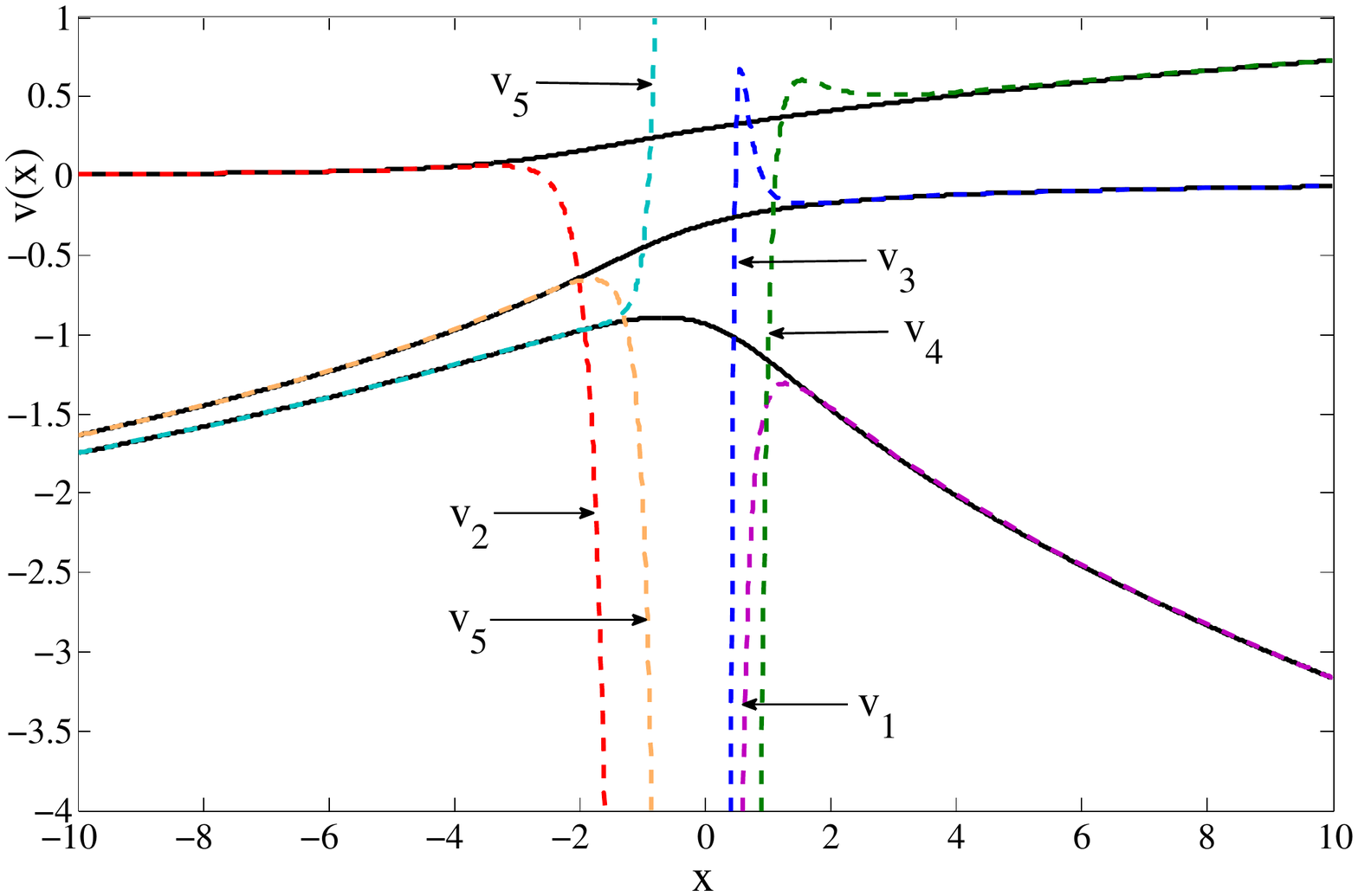}}
  \caption{\footnotesize{$n=4$ algebraic solutions with $\Gamma = 0$ and $c=1$.}
  }\label{n=4g0}
  \end{center}
\end{figure}

The $v_5|v_3$ solutions with $\Gamma = 0$ are new, similar to those above with
$c=0$. Further work is required to understand the existence and nature of such theories.

\item $c = \pm \Gamma$\\
Again we consider $c=\Gamma$ since the case of $c=-\Gamma$ is completely analogous.
Figure~\reef{n=4ceqg} plots the solutions for $c=1$ and $\Gamma =1$. The four smooth solutions 
which interpolate between $-x$ and $+x$ are
two $v_5|v_1$ solutions, one $v_5|v_4$ solution and one $v_5|v_3$ solution. (The two $v_5|v_1$ solutions
can be obtained from one another by applying $\IZ_2$ symmetries on the signs of the parameters).
All of these solutions are new, and it is possible that they correspond to well-defined
underlying string theories, but we leave any definitive claims to future work.

\begin{figure}[ht]
  \begin{center}
  \subfigure{
  \includegraphics[width=80mm, height = 60mm]{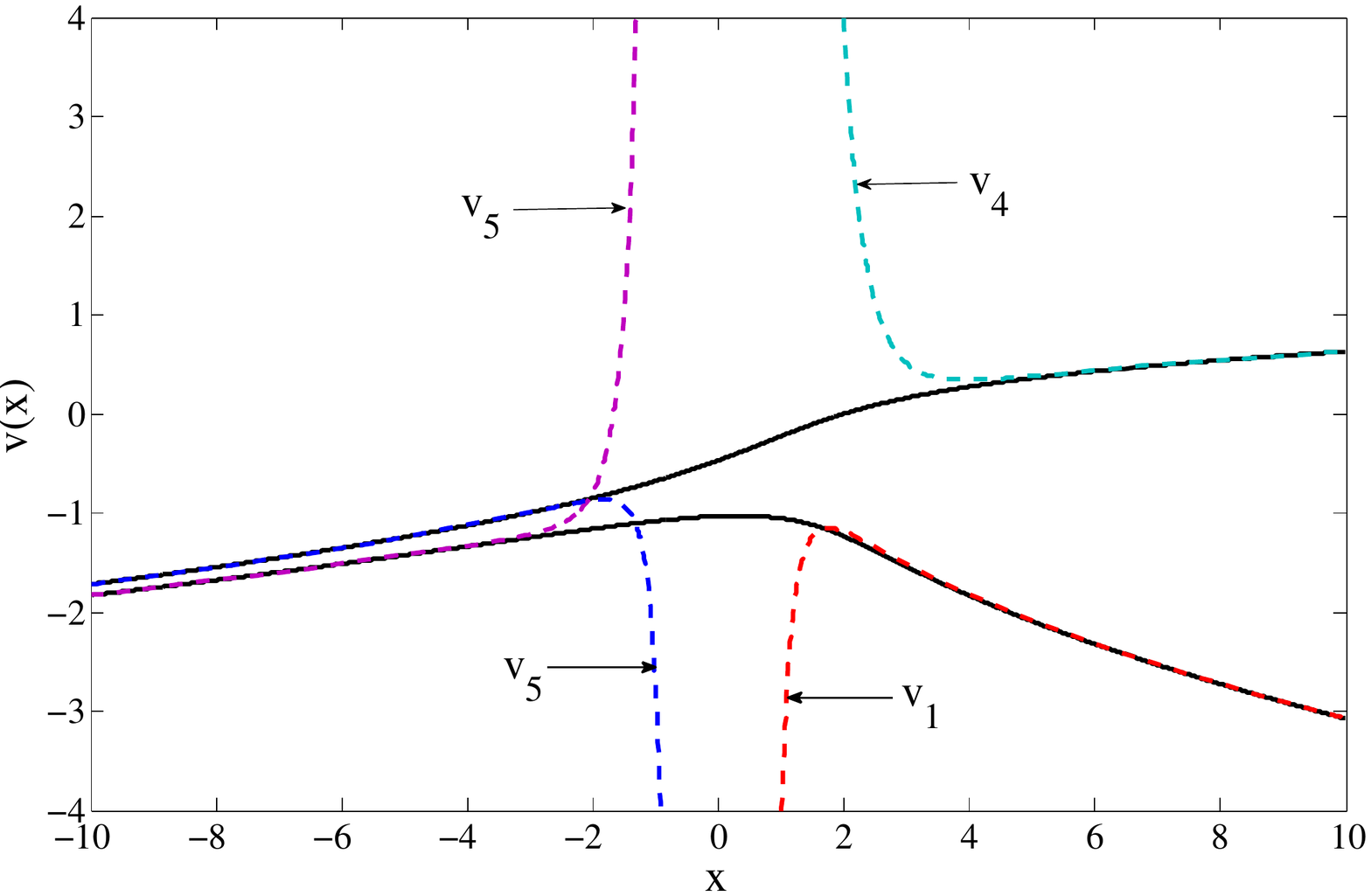}}
  \subfigure{\includegraphics[width=80mm, height = 60mm]{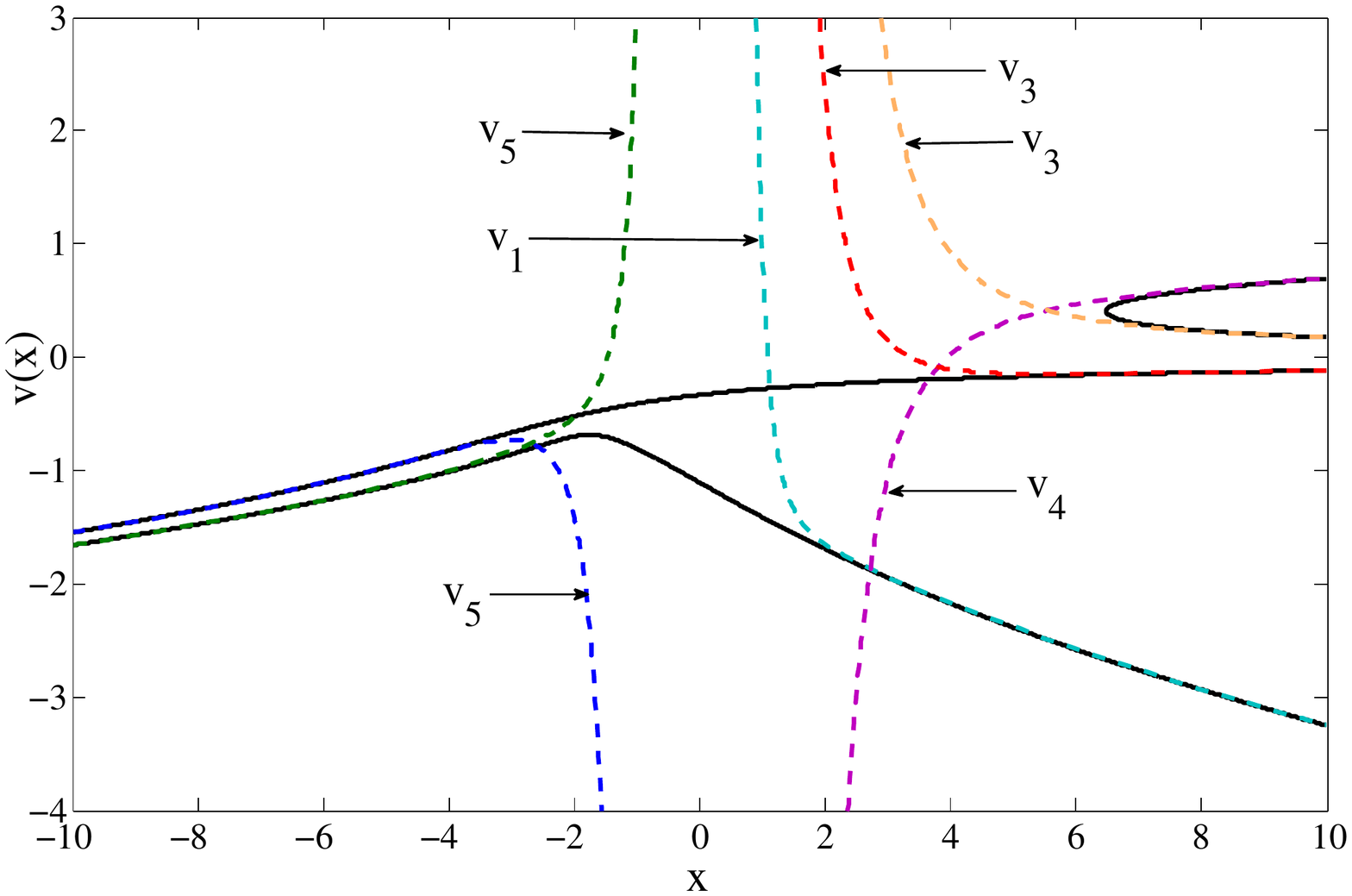}}
  \caption{\footnotesize{$n=4$ algebraic solutions with $c = 1$ and $\Gamma = 1$.}
  }\label{n=4ceqg}
  \end{center}
\end{figure}

\end{enumerate}

\section{DWW $n=1$: 't Hooft limit and Numerical
  Results}\label{sec:n=1expnmatch}
The string equations for $n=1$ are,
\begin{eqnarray}\label{DWWn=1}
2v - u_x + u^2 + g_1 x u = 2 \nu g_1 (c + \frac{1}{2}) \quad , \hspace{30mm} \\
\left(-v + \frac{1}{4}u^2 + \frac{1}{2}u_x\right)(u + g_1 x)^2 -\frac{1}{2}u_{xx}(u + g_1 x) + \frac{1}{4}\left(u_x + \nu g_1\right)^2  = \nu^2 g_1^2 \Gamma^2 \quad,  \nonumber
\end{eqnarray}
where we have used the relation $g_1 = \frac{1}{t_1}$. These equations are simple enough
to allow for complete numerical solutions, which will help us
classify new non-perturbatively complete solutions.

We will also examine these equations in the limit~\reef{modlimDWW},
under which they reduce to
\begin{eqnarray}\label{n=1algeqn}
u^2 + 2v -2 u x &=& -4 c \quad, \nonumber\\
v^2 + 2 v \left(u^2 -4 u x + 4 x^2 -4 c\right) &=& 4 (c^2 - \Gamma^2) \quad,
\end{eqnarray}
where we have used $g_1 = -2$. These equations admit four asymptotic expansions,
which we label as:
\begin{eqnarray}\label{lon=1}
v_2 &\sim& \frac{\nu^2}{x^2}\left(c^2 - \Gamma^2\right) \quad, \nonumber\\
v_{3a} &\sim& g_1 \nu \left(c + \Gamma\right) \quad, \nonumber\\
v_{3b} &\sim& g_1 \nu \left(c - \Gamma\right) \quad, \\
v_{4} &\sim& \frac{1}{9} g_1^2 x^2 \quad. \nonumber
\end{eqnarray}
The solutions with the three different parameter constraints are presented below.
\begin{enumerate}
\item $c = 0$ \\
 A plot of solutions to the algebraic equations~\reef{n=1algeqn} with $c = 0$
is shown in figure~\reef{c0gm2}. There are two smooth solutions interpolating
between $+x$ and $-x$, labeled $v_3|v_2$ in our convention.

\begin{figure}[!h]
  \begin{center}
  \includegraphics[width=70mm]{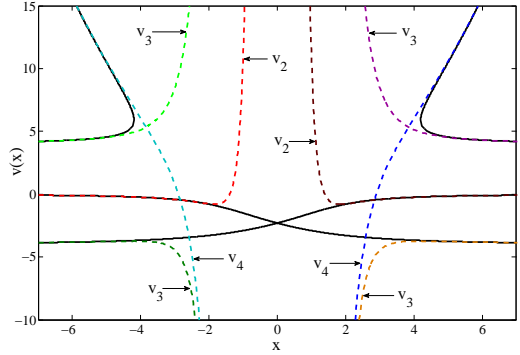}\\
  \caption{\footnotesize{$n=1$ algebraic solution with $c=0$ and $\Gamma = -2$.}
  }\label{c0gm2}
  \end{center}
\end{figure}

As outlined in the Appendix \reef{BraneFlux}, $v_2$ with $c = 0$ can
be interpreted as a flux expansion (for odd $n$) in $\Gamma$, while
$v_3$ with $c = 0$ is a brane expansion in $\Gamma$. These
interpretations are possible strictly for $c = 0$; a finite value of
$c$ results in powers of $g_s$ that do not allow $v_2$ to be
interpreted as a flux expansion.  So a string theory corresponding to
this solution (if it exists) should require $c=0$ in the exact string
equations.

\item $\Gamma = 0$ \\
  The plots for this case are shown in figure~\reef{cm2g0}. There are
  three smooth solutions, two of which are $v_4 | v_2$ and $v_2 | v_4$
  solutions. The third solution is parallel to the $x$--axis and
  exactly equals $v_{3a}$ and $v_{3b}$.  It is a 0B solution, the
  `topological point' of the 0B theory~\cite{Klebanov:2003wg} with $v
  = \nu g_1 c$.

  The $v_4 | v_2$ and $v_2|v_4$ solutions are new. For any value of
  $\Gamma$, $v_2$ and $v_4$ have powers of $c$ and $g_s$ consistent
  with a parameter counting branes.  This suggests that a string
  theory underlying such solutions (if it exists) should have branes
  in \emph{both} asymptotic regimes.

\begin{figure}[!h]
  \begin{center}
  \includegraphics[width=70mm]{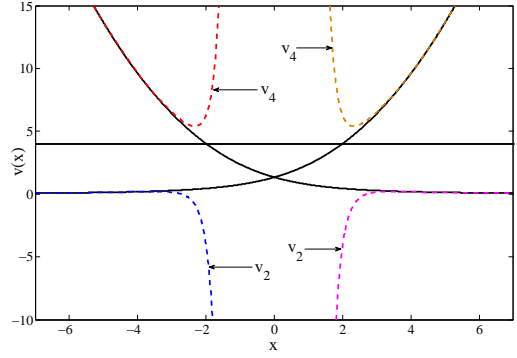}\\
  \caption{\footnotesize{$n=1$ algebraic solution with $c=-2$ and $\Gamma = 0$.}}\label{cm2g0}
  \end{center}
\end{figure}

Interestingly, algebraic $v_4 | v_2$ solutions do not exist for
$c>0$. Nevertheless, we have demonstrated the existence of numerical
solutions to the full equations in this and other cases. See
figure~\reef{cwithg0full}.  These solutions\footnote{Note the
  development of a bump in the interior of the solution as $c$
  decreases. There are a number of qualitative features of this family
  of solutions that are akin to those seen in studies of the type~0A
  case in refs.~\cite{Carlisle:2005mk,Carlisle:2005wa}, when examining
  the case of $\Gamma\to -1$.} were obtained using the ${\tt bvp4c}$
algorithm in $\textrm{MATLAB}$.  One lesson learned is that the
failure to find algebraic solutions in a t 'Hooft limit is not a
guarantee that smooth solutions to the full equations do not exist.

\begin{figure}[!h]
  \begin{center}
  \includegraphics[width=100mm]{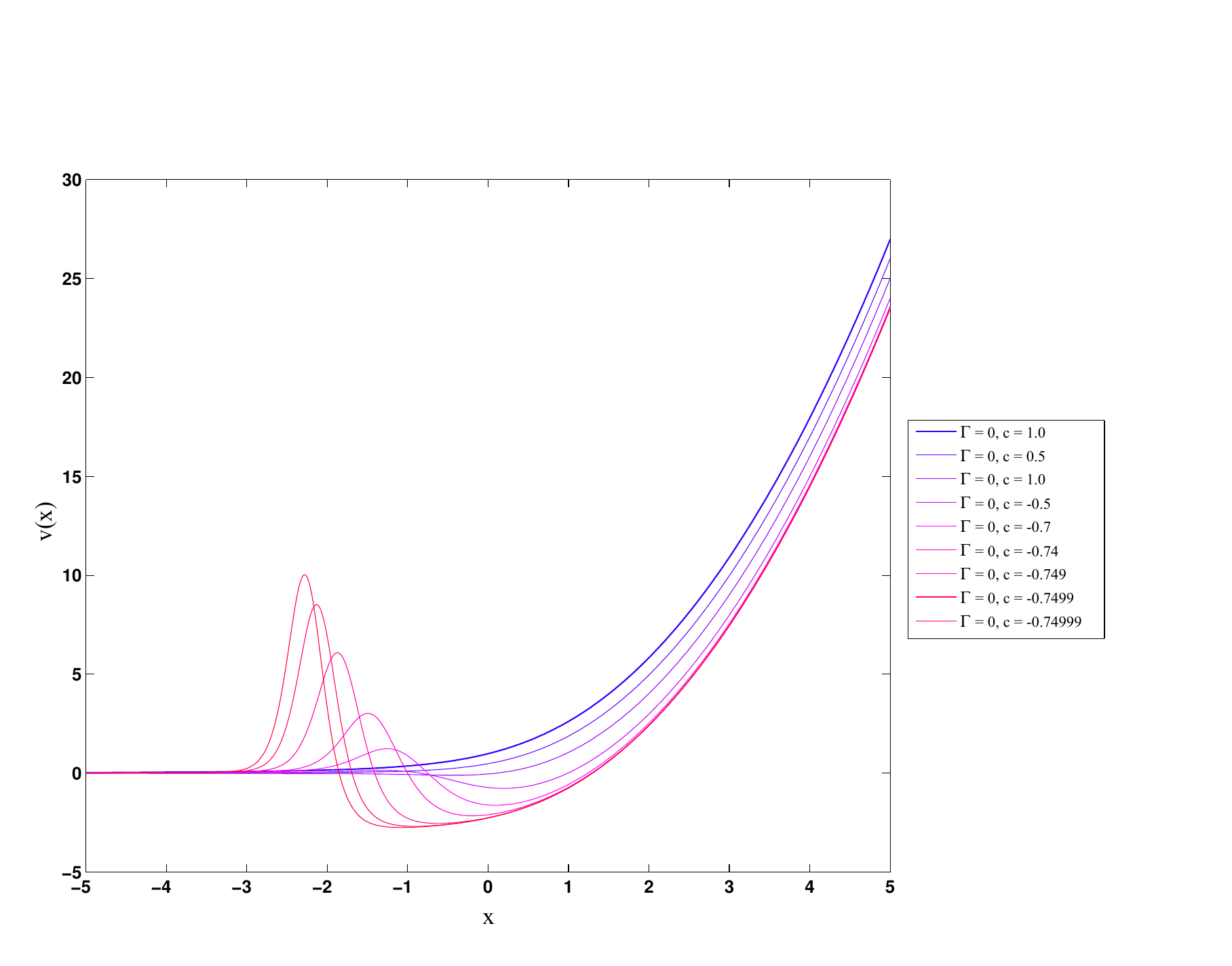}\\
  \caption{\footnotesize{$n=1$ solution to the full string equations~\reef{DWWn=1} with $\Gamma=0$.}
  }\label{cwithg0full}
  \end{center}
\end{figure}

\item $c = \pm \Gamma$\\
  Figure~\reef{c2g2} shows algebraic solutions where $c=2$ and
  $\Gamma=2$. These are $v_4|v_{3}$ and $v_{3}|v_4$ solutions in our
  notation.

  The constraint $c \pm \Gamma = 0$ on the algebraic equations lifts
  to $c \pm \Gamma = \xi$ on the full solutions. The constant $\xi$
  cannot be determined from algebraic solutions alone. In this case,
  however, a thorough numerical analysis of the full equations is
  possible and it shows that $\xi = - 1/2$. Examples of these
  numerical solutions are displayed in figure~\reef{cplusg}.

\begin{figure}[ht]
  \begin{center}
  \includegraphics[width=70mm]{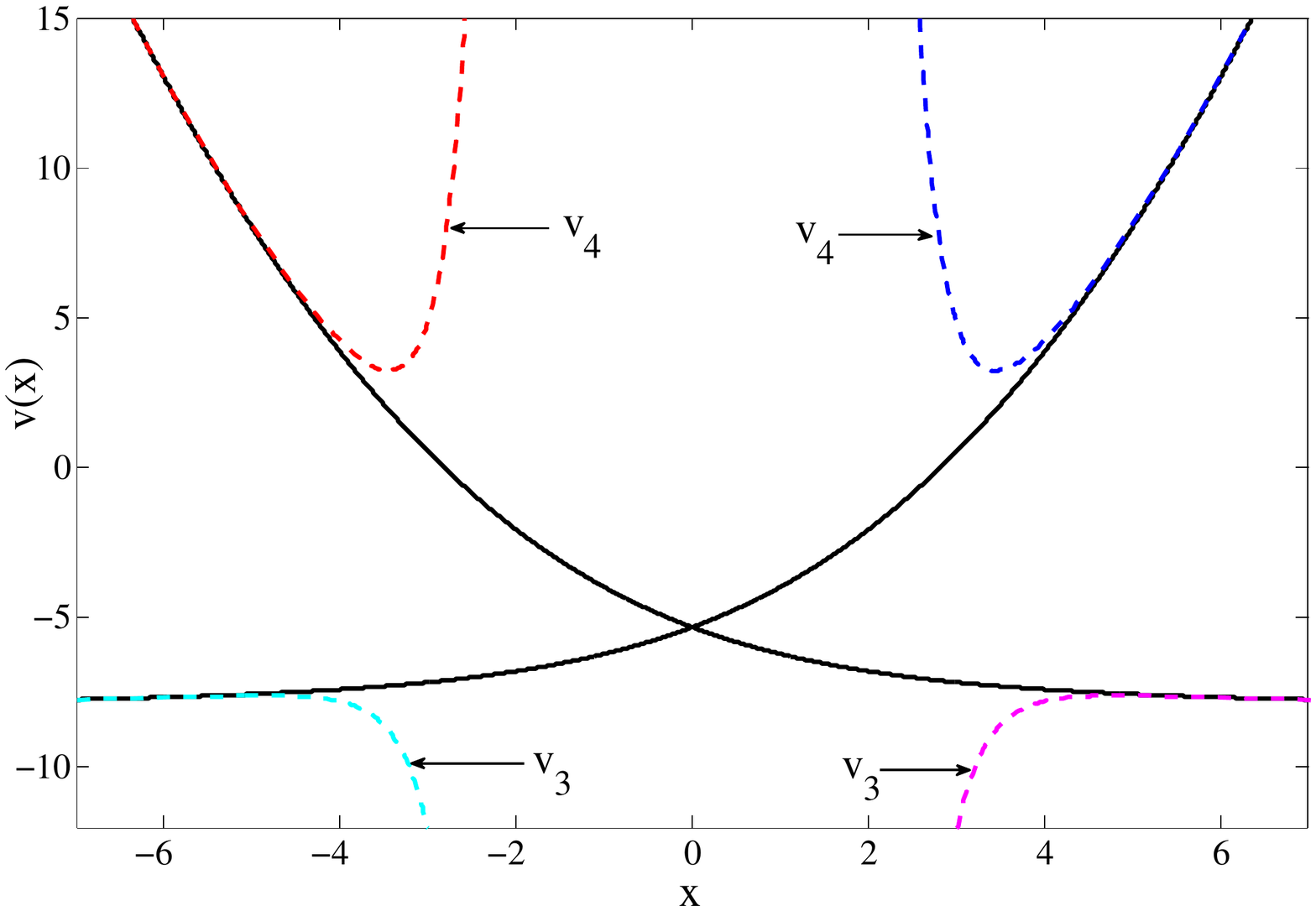}
  \caption{\footnotesize{$n=1$ algebraic solutions with $c = 2$ and $\Gamma= \pm2$.}
  }\label{c2g2}
  \end{center}
\end{figure}

\begin{figure}[ht]
  \begin{center}
  \includegraphics[width=100mm]{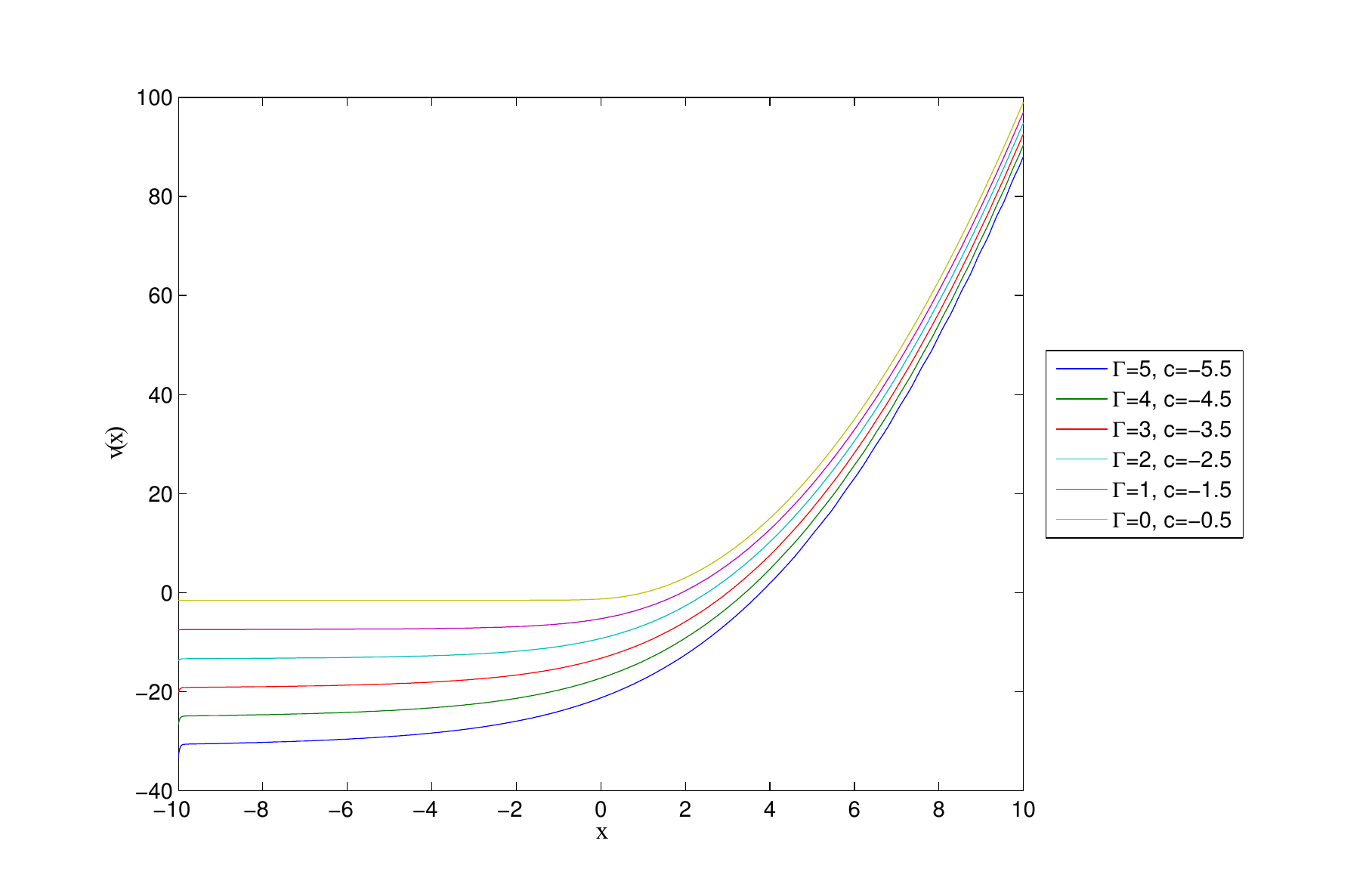}\\
  \caption{\footnotesize{$n=1$ solution to the full string equations~\reef{DWWn=1} with $c + \Gamma=-1/2$.}
  }\label{cplusg}
  \end{center}
\end{figure}

The numerical solution strongly suggests that such new solutions are
not just an artifact of the 't Hooft limit and should be taken
seriously as new examples of non--perturbatively complete solutions.
Such solutions were not apparent in the perturbative analysis
of ref.~\cite{DWW} and lead us to believe that the modified 't Hooft limit
presented here is a good way of unearthing new non--perturbative
solutions.

\end{enumerate}

\section{Discussion}
We have presented a modification of the 't Hooft limit, first used by
the authors of ref.~\cite{Klebanov:2003wg}, to argue for the existence
of new non--perturbative solutions to string equations that are
difficult to obtain numerically. We have analyzed the Painlev\'e~IV
hierarchy of string equations introduced in ref.~\cite{DWW} in this
limit, showing that examples of the conjectured type~II string
theories coupled to the $(4,4k-2)$ superconformal minimal models have
well--defined non--perturbative solutions. As in the case of type~0A,
higher $k$ solutions are likely to exist as a consequence of
these\cite{Johnson:1992pu}, since the underlying integrable flow
structure should evolve lower $k$ solutions into higher $k$ ones.

Although this limit results only in the highest powers of the
brane/flux parameter surviving (so that we have only spherical
topologies with boundaries or fluxes), it is likely that smooth
solutions exist for the full string equations too. The higher genus
terms in the free energy are obtained by including the derivative
corrections to the string equations and from the lesson of type~0A
(where we can compare to numerical solutions of the full equations ---
see section~\ref{sec:modlim}) it seems that they do not introduce any
singular behaviour.  This can presumably be checked through further
numerical work for the 0B and conjectured type~II theories.

We have uncovered a number of clear examples of new
non--perturbative solutions which also seem to be string--like. It would be
interesting to determine if these indeed correspond to new string
theories. This could presumably be checked using perturbative
techniques of the sort that we used in ref.~\cite{DWW} to identify the
type~II theories. We have demonstrated, in at least the $n=1$ case,
that such new solutions are not simply an artifact of the 't Hooft
limit, and that smooth solutions to the full equations can be obtained
numerically obeying the same parameter constraints. It would be
interesting to find new solutions of this type numerically for higher
$n$.

\section*{Acknowledgements}
This work was supported by the US Department of Energy.

\appendix
\section{The DWW expansions}\label{Appexpns}
We list the asymptotic expansions for $n=1$, $n=2$ and $n=4$ here. For
each $n$, there are multiple expansions within each class that can be
related to each other by various $\IZ_2$ symmetries. We list these
explicitly for $n=1$. The symmetries for $n=2$ and $n=4$ are similar
to those for $n=1$ and can be found in ref.~\cite{DWW}.

We also summarize the conditions under which a given expansion can be
thought of as encoding ZZ--branes or fluxes.

\subsection{$n = 1$}
\label{sec:expandone}
There are three classes of expansions for $v(x)$ (expansions in Class 1 appear only for even $n$),
\begin{eqnarray}\label{DWWn=1expn}
v_2&=&\frac{\nu^2}{x^2}(c^2-\Gamma^2)\left(1-\frac{\nu}{ g_1 x^2}6c+\frac{\nu^2}{g_1^2 x^4}(45c^2-5\Gamma^2+5)-\cdots\right) \quad ,\nonumber\\
v_3&=&\nu g_1(c-\Gamma)\left(1-\frac{\nu}{g_1 x^2}2\Gamma-\frac{\nu^2}{g_1^2 x^4}6\Gamma(c-3\Gamma)-\cdots\right) \quad ,\\
v_4&=&\frac{1}{9}g_1^2 x^2+\nu\frac{2 g_1 c}{3}-\frac{\nu^2}{x^2}\frac{1}{3}(3c^2+9\Gamma^2-1)+\frac{\nu^3}{g_1
x^4}6c(c^2-9\Gamma^2) - \cdots \quad .\nonumber
\end{eqnarray}

Integrating $v(x)$ twice, one can obtain the
free energy for a genus expansion of a
string theory which allows us to identify the string coupling to be
$g_{s} = {\nu}/{x^2}$.

\subsubsection{Symmetries for $n=1$}
The other expansions within each class can be obtained by the following
symmetry operation:
\begin{eqnarray*}
f_1: \Gamma \to -\Gamma \quad .
\end{eqnarray*}
Since $v_2$ and $v_4$ contain only even powers of $\Gamma$, the are
invariant under this map; however, $f_1\circ~v_3~\neq~v_3$. Thus there are
two expansions in the $v_3$ class, and, together with $v_2$ and $v_4$,
these comprise {four} total $n=1$ expansions.

\subsection{$n = 2$}
In this case, there are four relevant classes of expansions
\begin{eqnarray}\label{DWWn=2expn}
v_1&=&-g_2x-\frac{\nu g_2^{1/2}}{x^{1/2}}\Gamma+\frac{\nu^2}{x^2}\frac{1}{8}\left(-4c^2+4\Gamma^2+1\right) + \cdots \quad ,\nonumber\\
v_2&=&\frac{\nu^2}{x^2}\left(c^2-\Gamma^2\right)\left(1-\frac{2\nu^2}{g_2 x^3}(5c^2-\Gamma^2+1)+ \cdots \right) \quad ,\\
v_3&=&\frac{g_2^{1/2}\nu}{x^{1/2}}(c-\Gamma)\left(\frac{i}{\sqrt{2}}+\frac{\nu}{g_2^{1/2}x^{3/2}}\frac{1}{4}(c-5\Gamma)- \cdots\right) \quad ,\nonumber\\
v_4&=&-\frac{g_2x}{5}+\frac{\nu g_2^{1/2}}{x^{1/2}}\frac{ic}{\sqrt{5}}-\frac{\nu^2}{ x^2}\frac{1}{4}(2c^2+10\Gamma^2-1) -\cdots \quad .\nonumber
\end{eqnarray}
Here, we see that the string coupling is $g_{s} = {\nu}/{x^{\frac{3}{2}}}$.
There are $nine$ distinct expansions related to each other by various discrete symmetries
as for $n=1$.

\subsection{$n = 4$}
\label{sec:expandfour}
The following five classes of expansions are obtained:
\begin{eqnarray}\label{DWWn=4expn}
v_1(x) &=& -\frac{2 i }{\sqrt{3}}\sqrt{g_4}\sqrt{x} - \frac{\nu g_4^{1/4}}{x^{3/4}} \frac{(1+i)\Gamma }{2 \cdot 3^{1/4}} -\frac{\nu^2}{x^2}\frac{1}{48}\left(12 c^2-12\Gamma^2-5\right) + \cdots \quad ,\nonumber \\
v_2(x) &=&-\frac{\nu^2}{x^2}(\Gamma^2-c^2)\left(1-\frac{3\nu^4}{2 g_4 x^5}(21c^4-14 c^2 \Gamma^2 + \Gamma^4 + 35 c^2 - 5\Gamma^2 + 4)\right) + \cdots \quad ,\nonumber\\
v_3(x) &=& -\frac{g_4^{1/4} \nu}{x^{3/4}}(c-\Gamma)\left(\frac{1}{2^{3/4}(1+i)}+\frac{\nu}{g_4^{1/4} x^{5/4}}\frac{7\Gamma-3c}{8} + \cdots \right) \quad ,\\
v_4(x)&=&-\frac{2i}{3\sqrt{7}}\sqrt{g_4}\sqrt{x}-\frac{g_4^{1/4}\nu}{x^{3/4}}\frac{c}{\sqrt{3}\cdot7^{1/4}(1+i)}-\frac{\nu^2}{ x^2}\frac{1}{24}\left(6c^2+54\Gamma^2-5\right) + \cdots \quad ,\nonumber\\
v_5(x)&=&-2\sqrt{\frac{2}{21}}\sqrt{g_4}\sqrt{x}-\frac{g_4^{1/4}\nu}{x^{3/4}}\frac{21^{1/4}}{2^{5/4}}\left(-\frac{c}{\sqrt{7}}+\frac{\Gamma}{\sqrt{3}}\right)-\frac{\nu^2}{ x^2}\frac{1}{48}\left(12c^2-12\Gamma^2-5\right) + \cdots\nonumber \quad .
\end{eqnarray}
Here, the  string coupling is $g_{s} = {\nu}/{x^{\frac{5}{4}}}$.
There are a total of \emph{twenty five} expansions related by various $\IZ_2$ symmetries. These
have been explicitly listed in~\cite{DWW}.

\subsection{Brane/flux interpretation}\label{BraneFlux}
Recall the interpretation given to the parameter $\Gamma$ of the~0A
theory (and also to $q$ of the~0B theory). In the $\mu \rightarrow
+\infty$ regime, $\Gamma$ represents the number of background ZZ
D--branes in the model, with a factor of $\Gamma$ for each boundary in
the worldsheet expansion. Since an orientable surface with odd (even)
Euler characteristic must contain an odd (even) number of boundaries,
$\Gamma$ must be raised to an odd (even) power if $g_s$ is. In
addition, the power of $\Gamma$ must be less than or equal to the
power of $g_s$. In the $\mu \rightarrow -\infty$ regime, $\Gamma$
represents the number of units of R--R flux in the background, with
$g_s^2 \Gamma^2$ appearing when there is an insertion of pure R--R
flux. So in this case both $\Gamma$ and $g_s$ should appear with even
powers.

In applying these observations to the DWW expansions above, we
immediately notice the remarkable fact that the various expansions
have powers of the parameters which somehow allow for interpretations
as counting branes or fluxes.  This is by no means guaranteed, and
indeed its occurrence was one of our main motivations for in--depth
study of the system. The presence of two parameters, however, leads to
a few subtleties. For example, in some expansions an interpretation in
terms of branes is only possible if one of the two parameters is set
to zero. We summarize these below.

\subsubsection{Class 1}
\begin{itemize}
\item $v_1$ contains powers of $\Gamma$ consistent with a parameter
  counting branes. This remains true for any value of $c$.
\item $v_1$ contains powers of $c$ consistent with a parameter
  counting fluxes. However, for arbitrary $\Gamma$, $g_s$ appears with
  odd powers, inconsistent with our requirements for a description of
  fluxes, as mentioned above. This problem is avoided if we set
  $\Gamma=0$ since this forces the odd powers of $g_s$ to vanish.
\end{itemize}
\subsubsection{Class 2}
\begin{itemize}
\item For even $n$, $v_2$ contains powers of $\Gamma$ and $g_s$
  consistent with a parameter counting fluxes. This is true for all
  values of $c$. For odd $n$, the powers of $\Gamma$ are still
  consistent with the flux interpretation, but there are odd powers of
  $g_s$ which are inconsistent with fluxes. These odd powers can be
  removed by setting $c=0$.
\item For even $n$, $v_2$ contains powers of $c$ and $g_s$ consistent
  with a parameter counting fluxes. This is true for all values of
  $\Gamma$. For odd $n$, the powers of $c$ are consistent with a
  parameter counting branes. In this interpretation, there are no
  contributions from surfaces with only one boundary.
\end{itemize}
\subsubsection{Class 3}
\begin{itemize}
\item $v_3$ contains powers of $\Gamma$ consistent with a parameter
  counting branes. This is true only for $c=0$.
\item $v_3$ contains powers of $c$ consistent with a parameter
  counting branes. This is true only for $\Gamma=0$.
\item We notice that associating one boundary to each factor of $c$
  \emph{and} $\Gamma$ also produces a consistent worldsheet
  expansion. This encourages us to speculate these expansions might
  capture $c$ and $\Gamma$ simultaneously counting branes.
\end{itemize}
\subsubsection{Class 4}
\begin{itemize}
\item $v_4$ contains powers of $c$ consistent with a parameter
  counting branes. This remains true for any value of $\Gamma$.
\item $v_4$ contains powers of $\Gamma$ consistent with a parameter
  counting fluxes. However, for arbitrary $c$, $g_s$ appears with odd
  powers, inconsistent with our requirements for a description of
  fluxes, as mentioned above. This problem is avoided if we set $c=0$
  since this forces the odd powers of $g_s$ to vanish.
\end{itemize}
\subsubsection{Class 5}
\begin{itemize}
\item $v_5$ contains powers of $\Gamma$ consistent with a parameter counting branes. This is true only for $c=0$.
\item $v_5$ contains powers of $c$ consistent with a parameter counting branes. This is true only for $\Gamma=0$.
\item As for $v_3$, we notice that associating one boundary to each
  factor of $c$ \emph{and} $\Gamma$ also produces a consistent
  worldsheet expansion. This encourages us to speculate these
  expansions might capture $c$ and $\Gamma$ simultaneously counting
  branes.
\item These expansions do not exist for $n<3$.
\end{itemize}


\providecommand{\href}[2]{#2}\begingroup\raggedright\endgroup

\end{document}